\documentclass[times,sort&compress,3p]{elsarticle}
\usepackage[labelfont=bf]{caption}

\usepackage{amsmath,amsfonts,amssymb,amsthm,booktabs,color,epsfig,graphicx,hyperref,url}

\newcommand{\bA}{\mbox{\bf A}}

\newcommand{\bC}{\mbox{\bf C}}

\newcommand{\bD}{\mbox{\bf D}}
\newcommand{\bE}{\mbox{\bf E}}

\newcommand{\bI}{\mbox{\bf I}}

\newcommand{\bS}{\mbox{\bf S}}
\newcommand{\bU}{\mbox{\bf U}}
\newcommand{\bm}{\mbox{\bf m}}
\newcommand{\bM}{\mbox{\bf M}}
\newcommand{\bV}{\mbox{\bf V}}
\newcommand{\bW}{\mbox{\bf W}}
\newcommand{\bX}{\mbox{\bf X}}
\newcommand{\bY}{\mbox{\bf Y}}
\newcommand{\bZ}{\mbox{\bf Z}}
\newcommand{\bu}{\mbox{\bf u}}
\newcommand{\bv}{\mbox{\bf v}}
\newcommand{\bx}{\mbox{\bf x}}
\newcommand{\by}{\mbox{\bf y}}

\newcommand{\ba}{\mbox{\bf a}}
\newcommand{\bw}{\mbox{\bf w}}
\newcommand{\be}{\mbox{\bf e}}

\newcommand{\bzero}{\mbox{\bf 0}}

\newcommand{\bdelta}{\mbox{\boldmath $\delta$}}

\newcommand{\bleta}{\mbox{\boldmath $\eta$}}
\newcommand{\bmu}{\mbox{\boldmath $\mu$}}
\newcommand{\bSig}{\mbox{\boldmath $\Sigma$}}

\newcommand{\brho}{\mbox{\boldmath $\rho$}}

\newtheorem{theorem}{Theorem}[section]

\newtheorem{proposition}{Proposition}[section]
\newtheorem{lemma}{Lemma}%[section]

\newtheorem{condition}{Condition}
\usepackage[titlenumbered,linesnumbered,ruled,vlined]{algorithm2e}
\usepackage{multirow,ctable}

\begin{document}

\begin{frontmatter}

\title{Scalable Interpretable Learning for Multi-Response Error-in-Variables Regression}

\author[A1]{Jie Wu}
\author[A2]{Zemin Zheng}
\author[A3]{Yang Li}
\author[A4]{Yi Zhang}

\address[A1,A2,A3,A4]{The School of Management, University of Science and Technology of China, Hefei, Anhui 230026, P. R. China}
%\address[A1]{Address of Author Two in his country's language and rules}

%\cortext[mycorrespondingauthor]{Corresponding author. Email address: \url{zhengzm@ustc.edu.cn}}

\begin{abstract}
Corrupted data sets containing noisy or missing observations are prevalent in various contemporary applications such as economics, finance and bioinformatics. Despite the recent methodological and algorithmic advances in high-dimensional multi-response regression, how to achieve scalable and interpretable estimation under contaminated covariates is unclear. In this paper, we develop a new methodology called convex conditioned sequential sparse learning (COSS) for error-in-variables multi-response regression under both additive measurement errors and random missing data. It combines the strengths of the recently developed sequential sparse factor regression and the nearest positive semi-definite matrix projection, thus enjoying stepwise convexity and scalability in large-scale association analyses. Comprehensive theoretical guarantees are provided and we demonstrate the effectiveness of the proposed methodology through numerical studies.
\end{abstract}


\end{frontmatter}

\section{Introduction\label{sec:1}}

Large-scale association analysis is of great importance in many contemporary big data applications since it reveals hidden domain knowledge behind the data. For example, in genome-wide association studies, systematically estimating the genetic correlations between traits is crucial for understanding gene regulatory paths and gene functions, which provides insights into the genetic basis of quantitative variation in complex traits \citep{Baselmans2019}.
%identifying key genes is crucial for assessing the association between genetic markers and multivariate diffusion tensor imaging measurements, which offers greater power in inferring associations between genetic variants and various neurological or psychiatric illness \cite{wen2018}.
Similarly, in social network analyses, exploring the inter-dependency among users is a fundamental problem \citep{Kaiyang2017,zhu2019} and has interesting applications in crisis management \citep{starbird2012will} and spatial interactions analysis \citep{Kaiyang2019}.

Many powerful methods based on sparse reduced-rank regression were proposed to facilitate large-scale association network analysis. Specifically, the sequential estimation procedures proposed in \cite{mishra2017sequential,bahadori2016scalable} demonstrate scalability in large-scale applications by decomposing the estimation of the entire coefficient matrix into unit rank matrix recovery problems and thus guaranteed to stop in a few steps under low-rank structures. However, most of these existing methods are designed for clean data sets, while corrupted data are often encountered in various fields. Naively applying the aforementioned methods to analyze the corrupted data can lead to inconsistent and unstable estimates, thus drawing misleading conclusions. Therefore, it is urgent to develop scalable approaches for  high-dimensional multi-response regression under measurement errors.
%, especially in the high-dimensional settings \cite{liang2009variable,rosenbaum2010sparse}  such as engineering, economics, physics and biology

To alleviate the impacts of measurement errors, various statistical methods have been proposed. %for statistical learning in measurement error models
%\citep{CarollLiang1999,Carroll2010,liangCarrol2007,Carrol2011}.
Specifically, there is a line of work on dealing with measurement errors in univariate response linear regression models, which dates back to \cite{Bickel1987} and its extensions include \cite{liang2009variable,Lirunze2010}. Further development has been established in high-dimensional error-in-variables regression. For instance, \cite{stadler2012missing} developed an $\ell_1$-regularized likelihood approach to handle missing data by solving a negative log-likelihood optimization problem via EM algorithm. Similarly, \cite{loh2011high} proposed a Lasso-type estimator by replacing the corrupted Gram matrix with unbiased estimates for noisy or missing data. Furthermore, \cite{Belloni2017} developed a Dantzig selector-type estimator based on the compensated matrix uncertainty method. However, the negative likelihood functions are generally not convex after adjusting for the corrupted data and can depend on some crucial hidden parameters. To address this issue, \cite{datta2017} developed the convex conditioned Lasso (CoCoLasso) method by replacing the unbiased Gram matrix estimate with the nearest positive semi-definite matrix, thus enjoying the virtues of convex optimization and nice estimation accuracy in high-dimensional error-in-variables regression. The CoColasso was utilized as the initial estimate for statistical inference under sub-Gaussian designs when a fixed number of covariates are mentioned with errors \cite{Li2020}.

Despite the aforementioned progresses, there is relatively few work for high-dimensional error-in-variables regression with multivariate responses. A simple idea is to vectorize both the response matrix and the coefficient matrix such that the recently developed univariate response error-in-variables regression methods can be applied. However, it will ignore the multivariate nature of the correlated responses \citep{Izenman2008} and the appealing structures of the coefficient matrix such as low rankness and row sparsity.
%Since, we can apply univariate response error-in-variable regression method to them.
%For general structures of error-in-variables multivariate regression, to the best of our knowledge, there is no existing work which enjoys lower computation and fine estimation accuracy.
In this article, we develop a new approach called convex conditioned sequential sparse learning (COSS) to deal with multi-response regression under measurement errors by combining the strengths of the recently developed sequential sparse factor regression \citep{Zheng19} and the nearest positive semi-definite matrix projection \citep{datta2017}, thus enjoying stepwise convexity and scalability in large-scale association analyses.

The major contributions of this paper are threefold. First of all, the proposed method COSS is scalable and efficient in that it recovers the latent factors sequentially from the response matrix, which is not affected by the measurement errors of covariates. Second, we utilize the recently developed technique, the nearest positive semi-definite matrix projection, to alleviate the impacts of measurement errors when recovering the high-dimensional coefficient matrix from the latent factors. Thus, COSS enjoys stepwise convexity in view of either the regularization procedure or the regular eigenvalue decomposition, which renders it guaranteed computational stability. %which enjoyed high scalability for big data applications.
Last but not least, we provide comprehensive theoretical properties for the proposed method by establishing consistency in estimation, prediction, and rank selection. Numerical studies demonstrate the effectiveness of the proposed methodology.

The remainder of the article is organized as follows. Section \ref{sec2} presents the model setting and the new methodology. Theoretical properties including consistency in estimation, prediction, and rank selection are established in Section \ref{sec3}. We provide simulation examples in Section \ref{sec4}. Section \ref{sec5} concludes with extensions and possible future work. All technical details are relegated to the Supplementary Material.

\section{Multi-response regression under measurement errors}\label{sec2}

\subsection{Model setting}\label{sec2.1}

Consider a multi-response regression model
\begin{equation}\label{mod1}
\bY = \bX \bC^* + \bE,
\end{equation}
where $\bY = \{\by_1,\dots, \by_n \}^\top \in \mathbb{R}^{n \times q}$ is a multi-response matrix, $\bX = \{\bx_1,\dots, \bx_n \}^\top$ $ \in \mathbb{R}^{n \times p}$ is a fixed design matrix, $\bC^* \in \mathbb{R}^{p \times q}$ is an unknown coefficient matrix, and $\bE = \{ \be_1,\dots,\be_n\}^\top \in \mathbb{R}^{n \times q}$ is an error matrix with each row vectors $\be_i$ independent and identically distributed (i.i.d.) as $\mathcal{N} (\bzero,\bSig_E)$ \footnote{The Gaussian assumption can be relaxed as long as similar results to $\mathrm{E} \left(\| \mathbf{\Sigma}^{-1/2}_E \bE^\top \|_2 \right) \leq \sqrt{n} + \sqrt{q}$ can be ensured by the random matrix theory. See Lemma \ref{E_bound} for details.}. The columns of $\bX$ are assumed to have a common $\ell_2$-norm $\sqrt{n}$ and the matrix $\bC^*$ is assumed to be jointly low-rank and sparse.

To motivate our new method, we consider the regression coefficient matrix from a latent factor point of view similar to \cite{mishra2017sequential} and \cite{bahadori2016scalable}. Specifically, based on the SVD of $\bX \bC^*$, we have the following structure that
\begin{align}
\bX \bC^* & = (\bX \bU^*) \bD^* \bV^{*\top} = \bZ^* \bD^* \bV^{*\top}  , \label{eq:C*1} \\
         & \text{s.t.} \quad (\frac{1}{\sqrt{n}}\bX \bU^*)^\top (\frac{1}{\sqrt{n}}\bX \bU^*) = \bV^{*\top}\bV^* = \bI_{r^*}, \nonumber % \frac{1}{n}
\end{align}
where $\bC^* = \bU^* \bD^* \bV^{*\top} \in R^{p \times q}$, $\rm rank(\bC^*) = r^*$, $\bZ^* = \bX \bU^* = \{\bX \bu_1^*, \dots, \bX \bu^*_{r^*}\}$ $ \in R^{n \times r^*}$, $\bV^* = \{\bv^*_1, \dots, \bv^*_{r^*}\} \in R^{q \times r^*}$, $\bD^* = \text{diag}\{d_1^*, \dots, d^*_{r^*}\} \in R^{r^* \times r^*}$ is a diagonal matrix consisting of the singular values, and $\bI_{r^*}$ denotes the $r^* \times r^*$ identity matrix. In the above decomposition, $\bX \bU^*$ gives $r^*$ latent predictors/factors and $d_k^*\bv_k^{*}$ describes the strength and importance of the $k$th factor on the responses.

By rewriting the product $\bD^* \bV^{*\top}$ as a new matrix $\bV^{*\top}$ with the singular values in $\bD^*$ absorbed into the right singular vectors, we have
\begin{align}\label{SVD}
\bC^* = \sum_{k=1}^{r^*} \bu_k^* \bv_k^{*\top} = \sum_{k=1}^{r^*} \bC^*_k  = \bU^* \bV^{*\top},
\end{align}
where $\bC^*_k = \bu_k^* \bv_k^{*\top}$ is the $k$th layer unit rank matrix of $\bC^*$. Then $\bv_k^*$ in (\ref{SVD}) are no longer of unit length and the decomposition (\ref{SVD}) is a special one that gives uncorrelated latent factors $\bX \bu^*_k$ similarly as in factor analysis.

However, in many real applications, the design matrix we collect can contain unobserved measurement errors. In this paper, we consider two kinds of measurement errors associated with the design matrix $\bX$ listed as follows.

\begin{enumerate}
  \item[(1)] Additive errors. The observed covariates $\bW = \bX + \bA$, where the rows of the additive error matrix $\bA = (a_{ij})_{n \times q}$ are i.i.d. with mean vector $\bzero$ and covariance matrix $\bSig_A$.
  \item[(2)] Multiplicative errors. The observed covariates $\bW = \bX \odot \bM$, where $\odot$ denotes the Hadamard product and rows of $\bM = (m_{ij})_{n \times p}$, are i.i.d. with mean vector $\bmu_M$ and covariance matrix $\bSig_M$.
\end{enumerate}
Missing data can be viewed as a special case of multiplicative measurement errors with $m_{ij}$ = $I$($x_{ij}$ is not missing), where $I(\cdot)$ denotes the indicator function.

\subsection{Scalable estimation by COSS}
To get some insights of the proposed method, we first consider the noiseless case where $\bY^* = \bX \bC^*$ with $\bC^*$ adopting decomposition (\ref{SVD}). Explicitly, the latent factors $\bZ^*_k = \bX \bu^*_k, 1 \leq k \leq r^*$, are the top-$r^*$ eigenvectors of the regular eigenvalue problem
\begin{align*}%\label{true_step1}
\frac{1}{nq} \bY^* \bY^{*\top} \bZ = \lambda \bZ,
\end{align*}
corresponding to the eigenvalues $\lambda_k = \|\bX \bC^*_k\|_F^2 / nq$, and the right singular vectors $\bv_1^*, \dots, \bv_r^*$ can be written as
\begin{align*}%\label{v^star}
\bv_k^* = \frac{1}{\bZ_k^{*\top} \bZ_k^{*}} \bY^{*\top} \bZ_k^{*} = \frac{1}{n} \bY^{*\top} \bZ_k^{*}.
\end{align*}
We will make use of the fact that these two equations are not affected by the measurement errors as they mainly rely on the response $\bY^*$. Specifically, with corrupted data matrix $(\bW, \bY)$, we will recover the coefficient matrix $\bC^*$ sequentially in the following steps.

The first step is to solve the regular eigenvalue problem
\begin{align}\label{step1}
\frac{1}{nq} \bY \bY^{\top} \bZ = \lambda \bZ,
\end{align}
and get the estimated eigenvectors $\widehat{\bZ}_k$ with corresponding eigenvalues $\widehat{\lambda}_k$,
where $\widehat{\bZ}_k$ are assumed to have a common normalized $\ell_2$-norm $\sqrt{n}$, matching that of the true latent predictors $\bX \bu^*_k$. Then the right singular vectors $\bv_{k}^*$ can be estimated as
\begin{align*} %\label{v hat_rank}
\widehat{\bv}_k = \frac{1}{n} \bY^\top \widehat{\bZ}_k.
\end{align*}

The second step is to recover the sparse left singular vectors $\bu_k^*$ with data $(\widehat{\bZ}_k, \bW)$. Directly applying Lasso to the problem by minimizing
\begin{align*}
\frac{1}{2n} \|\widehat{\bZ}_k - \bW \bu \|_2^2 + &  \widetilde{\lambda}_k \| \bu \|_1 \Longleftrightarrow  \frac{1}{2} \bu^\top (\frac{1}{n} \bW^\top \bW) \bu - \frac{1}{n} \bu^\top \bW^\top \widehat{\bZ}_k + \widetilde{\lambda}_k \| \bu \|_1
\end{align*}
is often erroneous if the measurement errors are not ignorable \citep{rosenbaum2010sparse}. \cite{loh2011high} proposed to construct unbiased surrogates $\widehat{\bSig}$ and $\widetilde{\brho}^k$ for the unobservable $ \bSig$ and $\brho^k$ to alleviate the impacts of the measurement errors, in which $\bSig = \bX^\top \bX /n $ and $\brho^k = \bX^\top \widehat{\bZ}_k /n $. %(defined as $\frac{1}{n} \bX^T \bX$)(defined as $\frac{1}{n} \bX^T \widehat{\bZ}_k$)
Here the unbiased surrogates are defined as
\begin{align*}%\label{wucha}
\widehat{\bSig}_{\rm add} = n^{-1} \bW^\top \bW - \bSig_A, \widetilde{\brho}_{\rm add}^k = n^{-1} \bW^\top \widehat{\bZ}_k,
\end{align*}
for the additive errors setting or
\begin{align*}%\label{error_mul}
\widehat{\bSig}_{\rm mult} = n^{-1} \bW^\top  \bW \oslash (\bSig_M + \bmu_M \bmu_M^\top), \widetilde{\brho}_{\rm mult}^k = n^{-1} \bW^\top \widehat{\bZ}_k \oslash \bmu_M,
\end{align*}
for the multiplicative errors setting,
where $\oslash$ denotes the element-wise division operator for vectors and matrices. Similar to \cite{loh2011high,datta2017}, matrix $\bSig_A$ or ($\bSig_M$, $\bmu_M$) is assumed to be known for model identifiability \footnote{In practice, $\bSig_A$ or ($\bSig_M$, $\bmu_M$) can be obtained by repeated measurements or domain experience \citep{Carroll2010}.}.

However, the estimate $\widehat{\bSig}$ is generally not positive semi-definite in the high-dimensional setting such that the associated optimization problem can be no longer convex. To overcome the difficulties, we borrow the ideas from \cite{datta2017} to obtain the sparse left singular vectors $\widehat{\bu}_k$ by minimizing
\begin{align}\label{goal}
\widehat{\bu}_{k} = {\rm arg}\min_{\bu} \{ \frac{1}{2} \bu^\top \widetilde{\bSig} \bu - (\widetilde{\brho}^{\emph{k}})^\top  \bu + \widetilde{\lambda}_{\emph{k}} ||\bu||_1 \},
\end{align}
where $\widetilde{\lambda}_{\emph{k}}$ is the kth regularization parameter controlling sparsity, $\widetilde{\bSig}$ is a nearest positive semi-definite matrix defined as
% and can be tuned by cross-validation or certain information criterion
\begin{align}\label{semi-positive}
\widetilde{\bSig} = {\rm arg} \min_{\bSig \geq 0} ||\bSig - \widehat{\bSig}||_{max},
\end{align}
which can be efficiently solved by an alternating direction method of multipliers (ADMM). In fact, nonconvex regularization methods such as SCAD \cite{Fan2001} can also be applied to recover the singular vector $\bu_k^*$. Please refer to \cite{Zheng18} for the theoretical results of a general class of combined $\ell_1$ and concave penalties under measurement errors.

%Moreover, various kinds of regularization methods such as SCAD \cite{Fan2001} can also be applied in (\ref{goal}) to recover the singular vectors $\bu^*_k$ and a more complicated theory of the combined $\ell_1$ and concave regularization has already been established in \cite{Zheng18}.

%Moreover, various kinds of regularization methods such as SCAD \cite{Fan2001} and a more complicated theory of the combined $L_1$ and concave regularization \cite{Zheng18} can also be applied in (\ref{goal}) to recover the singular vectors $\bu^*_k$.

%In fact, the nearest positive semi-definite projection has already been generalized to a more complicated case of the combined $L_1$ and concave regularization  (Zheng et al.,2018).

Since the true rank $r^*$is unknown in practice, we can repeat these steps until the $k$th eigenvalue $\widehat{\lambda}_k$ of (\ref{step1}) is no larger than certain tolerance level $\mu$ \footnote{The tolerance level $\mu$ is set to be small such that all significant eigenvalues can be kept in the first step of COSS. In the numerical studies, we set $\mu = 1 \times 10^{-4}$ similarly as in \cite{bahadori2016scalable}.} and then tune the optimal rank $\widehat{r}$ by certain information criterion. Finally, the estimated coefficient matrix can be obtained by
\begin{equation*}%\label{C}
  \widehat{\bC} =\sum_{k = 1}^{\widehat{r}} \widehat{\bC}_k = \sum_{k=1}^{\widehat{r}}\widehat{\bu}_k \widehat{\bv}_k^{\top}.
\end{equation*}
The implementation of COSS is summarized in Algorithm \ref{alg:COCOSESS}.

\begin{algorithm}
\caption{COSS}
\label{alg:COCOSESS}

\SetAlgoLined
Input: $\bY \in \mathbb{R}^{n \times q}$, $\bW\in \mathbb{R}^{n \times p}$, $\bSig_A \in \mathbb{R}^{p \times p}$ or ($\bSig_M \in \mathbb{R}^{p \times p}$, $\bmu_M \in \mathbb{R}^{p \times 1}$) and a termination parameter $\mu$. \\
%\KwResult{how to write algorithm with \LaTeX2e }
set $k \gets 1$, $ j \gets 1$, $\widehat{\bY}_k \gets 0$ \\
$(\widehat{\bZ}_{k}, \widehat{\lambda}_k) \gets k$th eigenvector and eigenvalue of $(nq)^{-1}\bY \bY^\top \bZ = \lambda \bZ$ \\

\If{$\widehat{\lambda}_k  > \mu$}{
$\widehat{\bv}_k \gets \bY^\top \widehat{\bZ}_k /n$  \\
$\widehat{\bY}_{k} \gets \widehat{\bY}_{k} + \widehat{\bZ}_k \widehat{\bv}_k^\top$ \\
$k = k + 1$}

tune the optimal rank $\widehat{r}$ by information criterion (\ref{rank}) \\ %with sequences (k,$\widehat{\bY}_k$
$\widehat{\bSig}_{\rm add} = n^{-1} \bW^\top \bW - \bSig_A$ or $\widehat{\bSig}_{\rm mult} = n^{-1} \bW^\top  \bW \oslash (\bSig_M + \bmu_M \bmu_M^\top)$\\
$\widetilde{\bSig} = (\widehat{\bSig})_{+}$ \\ % and $\widetilde{\brho}^j = \widetilde{\brho}^j $

\While{$j  \leq \widehat{r}$}{
$\widetilde{\brho}_{\rm add}^j = n^{-1} \bW^\top \widehat{\bZ}_j$ or $ \widetilde{\brho}_{\rm mult}^j = n^{-1} \bW^\top \widehat{\bZ}_j \oslash \bmu_M $ \\
$\widehat{\bu}_j = {\rm arg}\min_{\bu} 2^{-1} \bu^\top \widetilde{\bSig} \bu - (\widetilde{\brho}^j)^\top \bu + \widetilde{\lambda}_j ||\bu||_1 $ \\
%$\widehat{\bv}_{k} \gets  \bZ_k^{T} \bW \widehat{\bu}_{k}/n$ \\
$\widehat{\bC}_j = \widehat{\bu}_{j}\widehat{\bv}_{j}^{\top}$ \\
$ j = j + 1$
}
$\widehat{\bC} = \sum_{j = 1}^{\widehat{r}} \widehat{\bC}_j$
\end{algorithm}

The proposed new method COSS enjoys two advantages. First, COSS takes full advantage of the fact that the response $\bY$ is not affected by measurement errors and thus can accurately recover the eigenvectors $\bZ^*_k$ in the first step. Second, by utilizing the nearest positive semi-definite matrix projection in the second step, we adjust for the measurement errors in a convex way when estimating the sparse left singular vectors. Therefore, COSS is stepwisely convex in view of either the penalization procedure or the regular eigenvalue decomposition, which renders it computational stability and estimation accuracy.

\section{Theoretical properties}\label{sec3}
In this section, we will list a few technical conditions and then analyze the theoretical properties of COSS.

\subsection{Technical conditions}
\begin{condition} \label{cond1}
The entries of measurement error matrices $\bA$ and $\bM$ are all independent and identically distributed sub-Gaussian random variables.
\end{condition}

\begin{condition} \label{cond2} There exist some constant $d_{\lambda}$ such that the top-$r^*$ population eigenvalues $\lambda_k$ satisfy $\lambda_{k} - \lambda_{k + 1} \geq d_{\lambda}$, $k = 1, \ldots, r^*$.
\end{condition}

\begin{condition}\label{cond3} The random error vectors $\be_i \in \mathbb{R}^{q}$ $ \sim \mathcal{N} (\bzero, \bSig_E)$ with the eigenvalues of matrix $\bSig_E$ bounded by positive constants $\gamma_u$ and $\gamma_l$ from above and below, respectively.
\end{condition}%[Bounded eigenvalues] There exists some such that

\begin{condition} \label{conduv} The $\ell_2$ norm of the left and right population singular vectors satisfy $\|\bu^*_k\|_2 \leq U$ and $\|\bv^*_k\|_2/\sqrt{q} \leq V$ for any $k$, $k = 1, \ldots, r^*$ with positive constants $U$ and $V$.
\end{condition}

\begin{condition}\label{cond5} The restricted eigenvalue of the Gram matrix $\bSig = \bX^\top \bX /n $ satisfy:
\begin{align*}
0<\Omega= \mathop{\min}\limits_{\bu \neq \bzero, \lVert \bu_{S^c} \rVert_1 \le 3 \lVert \bu_{S} \rVert_1}\; \frac{\bu^\top \bSig \bu}{ {\lVert \bu \rVert }_2^2},
\end{align*}
where $S$ is a support set of vector $\bu$ and $S^c$ denotes its complementary set.
\end{condition}
Condition \ref{cond1} puts a mild assumption on measurement error matrices since subgaussians are a natural kind of random variables for which the properties of Gaussians can be extended \cite{Buldygin1980}. Condition \ref{cond2} is essential for ensuring the identifiability of the latent factors $\bZ_k^*$. It requires distinct separation among successive nonzero eigenvalues $\lambda_k$ such that the latent factors are distinguishable. Similar conditions can be found in \cite{bahadori2016scalable,Uematsu17}.

Condition \ref{cond3} imposes upper and lower bounds on the eigenvalues of noise covariance matrix $\bSig_E$, which is weaker than the regular assumptions on the noise vectors utilized in \cite{bunea2012joint,liu2015calibrated,Uematsu17} since they need strict diagonal structure of the noise covariance matrix.

Condition \ref{conduv} is imposed to give upper bounds on the lengths of the left and right population singular vectors. The magnitudes of left singular vectors $\|\bu^*_k\|_2$ are $O(1)$, which is reasonable as $\|\bX \bu^*_k \|_2 $ are of the magnitude of $\sqrt{n}$ given in (\ref{eq:C*1}). Besides, $\bv^*_k$ are $q$-dimensional vectors such that there is an extra factor $1/\sqrt{q}$ in the second inequality.
Condition \ref{cond5} is the restricted eigenvalue condition proposed in \cite{Ritov2009}, and is widely used in lasso related articles. %, see, for example, \cite{datta2017,Wadsworth2014Asymptotic,van2009}.
It imposes a lower bound on eigenvalues of the Gram matrix $\bSig$ to constrain the correlations between relatively small numbers of predictors in the design matrix $\bX$.

\subsection{Main results}
\begin{proposition}\label{rho}
Assume that the distribution of $\widehat{\bSig}$ and $\widetilde{\brho}^k$ are identified by a set of parameters $\theta$. Then there exist universal constants C and c, and positive functions $\zeta$ and $\epsilon_0$ depending on $\theta$ and $\sigma^2$ such that for any $\epsilon \leq \epsilon_0$, $\widehat{\bSig}$ and $\widetilde{\brho}^k$ satisfy the following probability statements:
\begin{align}\label{con:sig}
& \Pr ( \mid\widehat{\bSig}_{ij} - \bSig_{ij}\mid\geq\epsilon)\le C \exp(-cn\epsilon^2\zeta^{-1}) \quad \forall i,j = \{1,\dots, p\}\\ \label{con:rho}
& \Pr ( \mid\widetilde{\brho}^k_{j} - \brho^k_{j}\mid\geq\epsilon)\le C \exp(-cn\epsilon^2s^{-2} \zeta^{-1}) \quad  \forall j = \{1,\dots,p\};    k = \{1,\dots, r^*\}.
\end{align}
\end{proposition}
Proposition \ref{rho} shows that the surrogates $\widehat{\bSig}$ (and hence $\widetilde{\bSig}$) and $\widetilde{\brho}^k$ can be sufficiently close to $\bSig$ and $\brho^k$ respectively in terms of the element-wise maximum norm. The rates of the tail probability bounds for $\widehat{\bSig}$ and $\widetilde{\brho}^k$ are the same as those in \cite{datta2017}, which were established for univariate response linear regression models with sub-Gaussian distributed measurements errors. Since the corrupted design matrix $\bW$ is exactly the same as that in \cite{datta2017}, inequality (\ref{con:sig}) holds automatically under Condition \ref{cond1}. Our main contribution is equality (\ref{con:rho}), which shows that $\widetilde{\brho}^k$ satisfy the same concentration inequality even if the response vector is replaced with the estimated latent factors $\widehat{\bZ}_k$. In view of the results, even if $\bSig_A$ or $(\bSig_M, \bmu_M)$ is replaced with some estimates, our theories are still valid as long as the gap between them is no more than the magnitude of $\epsilon$.

\begin{theorem}[Consistency of sequential estimation]\label{theo1}
Assume that the Conditions 1-5 hold, and $\max\{s \sqrt{(\zeta \log p)/n}$, $\widetilde{C}(\sqrt{n} + \sqrt{q})/ \sqrt{nq} \} < \widetilde{\lambda}_k \leq \min (6 \epsilon_0, 12 \epsilon_0 B )$ with $\widetilde{C} > 24\sqrt{\lambda_1}\gamma_u/d_{\lambda}$ and $B = \max_{1 \leq k \leq r^*} \|\bu^*_k \|_{\infty}$. Then for sufficiently large $n$ and any $k$, $1\leq k \leq r^*$, the following statements hold with probability at least $1 - C p^{-c} $, % $1- C \exp ( - c\log p )$,
\begin{align*}
\|\widehat{\bu}_{k} - \bu_{k}^* \|_{2}  &\leq C_{u}\sqrt{s} \widetilde{\lambda}_k ,\
\frac{1}{\sqrt{q}} \|\widehat{\bC}_k - \bC^*_k \|_F  \leq (V C_{u} + U C_{v}) \sqrt{s} \widetilde{\lambda}_k,  \\
\frac{1}{\sqrt{n}} \|\bX\widehat{\bu}_{k} - \bX \bu^*_{k}\|_2  &\leq \widetilde{C}_{u}\sqrt{s} \widetilde{\lambda}_k, \
\frac{1}{\sqrt{nq}}\|\bX \widehat{\bC}_k - \bX  \bC^*_k \|_F   \leq (V \widetilde{C}_{u} + C_{v}) \sqrt{s} \widetilde{\lambda}_k,
\end{align*}
where $s = \max_{k = 1}^{r^*} |S_k|$ is the maximum sparsity level with $S_k$ the support set of the true left singular vector $\bu^*_k$, and $C$, $c$, $C_{u} = 4 \sqrt{2} /\Omega$,  $\widetilde{C}_{u} = 4 \sqrt{2}/ \sqrt{\Omega}$, and $C_{v} = \sqrt{\lambda_1} \cdot \widetilde{C}+ 2\gamma_u$ are positive constants.
\end{theorem}
Theorem 3.1 establishes the estimation error bounds for top-$r^*$ singular vectors $\bu_k^*$, latent factors $ (\bX \bu^*_k) /\sqrt{n}$ and unit rank matrices $\bC^*_k / \sqrt{q}$, and the prediction error bounds for top-$r^*$ layer matrices $\bX \bC^*_k / \sqrt{nq}$. From this theorem, we see that the convergence rates of these bounds are all in the same order of $O(\sqrt{s} \widetilde{\lambda}_k)$ with significant probability. Since $s \sqrt{\zeta(\log p)/n}$ is typically larger than $(\sqrt{n} + \sqrt{q})/ \sqrt{nq}$ when $q$ is of high dimensionality, the convergence rates here are slower than those established under the clean data setting by a factor of $s$, in view of the magnitude of the regularization parameter $\widetilde{\lambda}_k$, which is similar to the univariate response setting in \cite{datta2017}.
%From this theorem, we see that the dimension constraint for the consistency is $s^2\log p\ll n$ as the dimensionality $q$ is around the same level as the sample size $n$. Moreover, the convergence rates of these bounds are all in the same order of $O(\sqrt{s} \widetilde{\lambda}_k)$ with significant probability.
%A distinctive feature is that the tuning parameter $\widetilde{\lambda}_k$ is of different magnitude from that in \cite{datta2017} to suppress the noise in the univariate response error-in-variables regression for recovering the sparse left singular vector $\bu_{k}^*$. Furthermore, it is worth pointing out that the convergence rates are generally slower than $O( \sqrt{s}(\sqrt{n} + \sqrt{q})/\sqrt{nq})$ established in \citep{Zheng19} under the clean data setting, which coincides with the price in accuracy we pay to trade for the corrupted data and is consistent with the results in \cite{datta2017}.
%Therefore, our proposed method COSS not only has a significant advantage in computational efficiency, but also enjoys appealing asymptotic properties.

Based on the discussion before, low-rank coefficient matrix $\bC^*$ can be accurately recovered as long as the rank is correctly identified. In particular, we propose the following BIC-type information criterion to tune the true rank.

\begin{theorem}[Consistency of rank recovery]\label{theo2}
Suppose that Conditions \ref{cond2}-\ref{conduv} hold, $r \{(\sqrt{n} + \sqrt{q})/\sqrt{nq}\}^{1/2} = o(1)$, ${r^*} $ $(\log n/\sqrt{n})^{1/2} = o(1)$, and $\sqrt{n}/(\sqrt{q} \log n) = o (1)$. Then for sufficiently large $n$, the information criterion is defined as
\begin{align}\label{rank}
\mathcal{C}({k}) = \sqrt{n} \log \mathcal{L}(k) + k \log n,
\end{align}
where $\mathcal{L}(k) = \|\bY - \widehat{\bY}_{k}\|_F^2 / nq$ with $\widehat{\bY}_{k} = \sum_{j=1}^{k} \widehat{\bZ}_j \widehat{\bv}_j^\top$. It attains its minimum value when $k = r^*$ with probability at least $1 - c_0 \exp(-n/2)$ for some positive constant $c_0$.
\end{theorem}

Theorem \ref{theo2} proposes a BIC-type information criterion (\ref{rank}) to consistently identify the true rank $r^*$. In fact, since both the latent factors $\widehat{\bZ}_j$ and the right singular vectors $\widehat{\bv}_j$ only depend on the response $\bY$, the tuning of rank will not be affected by the corrupted data by utilizing the proposed method. As the multi-response regression is decomposed into several univariate response regressions in the first step, the optimal sparsity parameters $\widetilde{\lambda}_k$ in (\ref{goal}) can be tuned by cross validation or certain information criterion separately.

We have shown that the proposed method COSS can enjoy the appealing asymptotic properties. However, the aforementioned theoretical results are based on the assumption that the covariance matrix of measurement errors is known. When $\bSig_A$ or $(\bSig_M, \bmu_M)$ is unknown and must be estimated from the data, similar results can still hold by repeated measurements \cite{loh2011high}. %(Carroll et al., 1995;  Loh and Wainwright, 2012).
Moreover, the fixed design is not essential,  and similar theoretical conclusions can also be extend to sub-Gaussian case by using the same argument as that in \cite{Li2020}. %Li et al. (2020).

\section{Simulation studies}\label{sec4}
In this section, we investigate the finite-sample performance of the proposed method COSS. Two methods designed for clean data sets are employed for comparison to illustrate the impacts of measurement errors if ignoring them. One is the rank constrained group lasso (RCGL) \citep{bunea2012joint} and the other is the sequential estimation with eigen-decomposition (SEED) \citep{bahadori2016scalable}. The three methods were implemented as follows. RCGL was implemented by the R package `\textsf{rrpack}' with the regularization parameter tuned by BIC and the rank tuned by the criterion of joint rank and row selection (JRRS), as suggested in \cite{bunea2012joint}. SEED selected the tuning parameters by minimizing the surrogate prediction error calculated based on an independent validation set with its size equal to the sample size, similarly as in \cite{bahadori2016scalable}. COSS utilized the R package `\textsf{lars}' for sparse regression with the regularization parameter tuned by BIC and the rank tuned by information criterion \eqref{rank}. Both additive and multiplicative measurement errors are included in the studies, as well as the missing data case.
%In this section, we proceed with investigating the finite-sample performance of the proposed method COSS. By contrast, we utilized rank constrained group lasso (RCGL) by \cite{bunea2012joint} and sequential estimation with eigen-decomposition (SEED) by \cite{bahadori2016scalable} to observe the impacts of measurement errors in that neither of them accounts for the corrupted data. Among them, RCGL was implemented by R package \textsf{rrpack} with regularization parameter tuned by BIC and rank tuned by JRRS, as suggested in \cite{bunea2012joint}. SEED selected termination and sparsity thresholds by minimizing the surrogate prediction error calculated based on an independent validation set with size equal to the sample size, as suggested in \cite{bahadori2016scalable}. COSS was implemented by utilizing the R package \textsf{lars} for univariate response regression with sparsity parameter tuned by BIC and rank tuned by BIC-type information criterion. Additive, multiplicative measurement errors and missing data are considered in the simulation studies. %Missing data as a special case was analyzed in a new simulation study in subsection \ref{simu2}.

%\subsection{Simulation example 1}\label{simu1}

We generated 100 data sets from multi-response regression model (\ref{mod1}). For each data set, the rows of $\bX \in \mathbb{R}^{n \times p}$ were independent and identically distributed (i.i.d.) vectors from $ \mathcal{N} (\bzero,\bSig_X)$ with $\bSig_X=(0.5^{|i-j|})_{p \times p}$. Then based on different types of measurement errors, the corrupted covariates matrices $\bW$ can be provided respectively as follows.

\textbf{Additive errors case.} The observed covariates $\bW = \bX + \bA$, where the rows of $\bA$ are i.i.d. vectors copied from $ \mathcal{N} (\bzero, \tau^2 \bI)$ with $\tau=$ 0.2.

\textbf{Multiplicative errors case.} The observed covariates $\bW = \bX \odot \bM$, where the components of $\bM = ((m_{ij}))$ follow log-normal distribution, meaning that $\log(m_{ij})$'s are i.i.d. variables from $ \mathcal{N} (0,\tau^2)$ with $\tau=$ 0.2.

\textbf{Missing data case.} The observed covariates are defined as $w_{ij} = x_{ij} m_{ij}$, $m_{ij} = I$($x_{ij}$ is not missing), so that the covariates are missing at random with probability 0.1. %with $pr = 0.05$

Similarly, the rows of noise matrix $\bE$ were i.i.d. random vectors from $ \mathcal{N}(\bzero, \gamma \bSig_E)$ with $\bSig_E=(0.5^{|i-j|})_{q \times q}$ and $\gamma = 0.1$. The generation of coefficient matrix $\bC^*$ was shown as follows. After creating matrix $\bC \in \mathbb{R}^{p \times q}$ with about 90 non-zero entries, each of which was i.i.d. from $ \mathcal{N}(0,1)$, we find the top-$r$ singular value decomposition of $\bC$ as $\bC = \bU \bS \bV^T$. Then we reparameterized the matrix $\bS$ by replacing the first $r$ diagonal entries of it by 100,99,...,100-$r$ and the rest by 0. Here we consider settings with $(n,q,r) = (200,300,10)$, whose dimensionality $p$ varies in $\{ 200,400,600,800\}$.

\begin{table}%[h]
\centering
\caption{Simulation Results \label{t1}}

    \smallskip
    \smallskip
	\centering
	\begin{tabular}{@{}p{0.5cm}c c c c c}
		\hline
		\multirow{2}{*}{$p$} & & Normalized  & Normalized  &  Rank\\
\smallskip
		                     & Algorithm & Prediction & Estimation & Recovery \\
\smallskip
		                     & & Error($\times 10^{-2}$)  & Error($\times 10^{-2}$)  & Error \\
		\hline
		\multicolumn{5}{c}{Additive error case }\\
		%\toprule
		%$p$ & Algorithm & Normalized Prediction Error & Normalized Estimation Error & Support Recovery AUC & Rank Recovery Error\\
		\hline
		\multirow{3}{*}{$200$}
		%&SESS$^*$ & $1.87~(0.00)$ & $0.94~(0.00)$ \\
		& SEED & $27.74~(0.12)$ & $22.27~(0.13)$ & $ 3.70~(0.30)$  \\
        & RCGL & $19.98~(0.22)$ & $21.55~(0.21)$ & $0.00~(0.00)$\\
		& COSS & $10.67 ~(0.05)$ & $11.14~(0.6)$ & $ 0.00~(0.00)$  \\
		\hline
		\multirow{3}{*}{$400$}
		%&SESS$^*$ & $1.87~(0.00)$ & $0.76~(0.00)$ \\
		& SEED & $21.09~(0.34)$ & $24.54~(0.58)$ & $ 4.06~(0.33)$ \\
        & RCGL & $25.78~(0.15)$ & $28.56~(0.21)$ & $ 0.00~(0.00)$\\
		& COSS & $10.44 ~(0.06)$ & $11.68 ~(0.06)$ & $ 0.00~(0.00)$  \\
		\hline
		\multirow{3}{*}{$600$}
		%&SESS$^*$ & $1.87~(0.00)$ & $0.78~(0.00)$ \\
		& SEED & $28.86~(0.32)$ & $32.90~(0.50)$ & $4.20~(0.22)$ \\
        & RCGL & $21.11~(0.20)$ & $22.37~(0.27)$ & $0.00~(0.00)$\\
		& COSS & $ 10.91~(0.05)$ & $ 11.27~(0.06)$ & $ 0.00~(0.00)$ \\
		\hline
        \multirow{3}{*}{$800$}
        %&SESS$^*$ & $1.86~(0.00)$ & $0.75~(0.00)$ \\
		& SEED & $32.28~(0.24)$ & $33.67~(0.36)$ & $4.02~(0.00)$ \\
        & RCGL & $28.97~(0.16)$ & $29.79~(0.23)$ & $0.00~(0.00)$ \\
        & COSS & $11.45~(0.07)$ &$12.77~(0.08)$ & $0.00~(0.00)$  \\
        \hline

		\multicolumn{5}{c}{Multiplicative error case}\\

	    \hline
		\multirow{3}{*}{$200$}
		%&SESS$^*$ & $1.87~(0.00)$ & $0.94~(0.00)$ \\
		& SEED & $26.76~(0.78)$ & $26.82~(0.82)$  & $3.54~(0.20)$  \\
        & RCGL & $16.28~(0.76)$ & $16.04~(0.78)$  & $0.00~(0.00)$\\
		& COSS & $ 10.64~(0.40)$ & $10.54~(0.41)$   & $0.00~(0.00)$  \\
		\hline
		\multirow{3}{*}{$400$}
		%&SESS$^*$ & $1.87~(0.00)$ & $0.76~(0.00)$ \\
		& SEED & $26.08~(0.75)$ & $25.99~(0.82)$  & $2.20~(0.43)$ \\
        & RCGL & $16.57~(1.00)$ & $16.29~(1.01)$  & $0.00~(0.00)$\\
		& COSS & $11.96~(0.35)$ & $12.38~(0.35)$  & $0.00~(0.00)$  \\
		\hline
		\multirow{3}{*}{$600$}
		%&SESS$^*$ & $1.87~(0.00)$ & $0.78~(0.00)$ \\
		& SEED & $29.13~(0.64)$ & $32.06~(0.74)$  & $3.40~(0.33)$ \\
        & RCGL & $17.16~(0.88)$ & $16.93~(0.89)$  & $0.00~(0.00)$ \\
		& COSS & $10.24~(0.35)$ & $10.25~(0.36)$  & $0.00~(0.00)$  \\
		\hline
        \multirow{3}{*}{$800$}
		%&SESS$^*$ & $1.88~(0.00)$ & $0.75~(0.00)$ \\
		& SEED & $31.88~(0.81)$ & $34.55~(0.81)$  & $3.20~(0.31)$ \\
        & RCGL & $18.78~(0.42)$ & $20.49~(0.51)$  & $0.00~(0.00)$\\
        & COSS & $12.56~(0.40)$ &$12.41~(0.42)$   & $0.00~(0.00)$ \\
        \hline
        \multicolumn{5}{c}{Missing data case}\\

\hline
\multirow{3}{*}{$200$}
          &SEED&26.71~(0.37)&27.74~(0.48)& 1.53~(0.15)   \\
          &RCGL&20.45(0.44)&23.86~(0.45)& 0.00~(0.00)   \\
          &COSS&9.90~(0.28)&9.43~(0.29)& 0.00~(0.00)     \\ \hline
		\multirow{3}{*}{$400$}
          &SEED&27.62~(0.73)&29.63~(0.98)&1.62~(0.13)    \\
          &RCGL&20.93~(0.45)&22.10~(0.50)&0.00~(0.00)   \\
          &COSS&9.13~(0.20)&8.86~(0.20)&0.00~(0.00)  \\\hline
          \multirow{3}{*}{$600$}
          &SEED&33.88~(1.27)&38.48~(1.32)&2.53~(0.35)        \\
          &RCGL&21.17~(0.47)&20.52~(0.50)&0.00~(0.00)    \\
          &COSS&11.86~(0.30)&11.92~(0.40)&0.00~(0.00)    \\\hline
          \multirow{3}{*}{$800$}
          &SEED&34.52~(1.33)&35.21~(1.25)&3.12~(0.36)        \\
          &RCGL&22.51~(0.14)&23.24~(0.25)&0.00~(0.00)   \\
          &COSS&13.30~(0.03)&13.55~(0.03)&0.00~(0.00)     \\\hline
  	
%		\multirow{3}{*}{$200$}
%		%&SESS$^*$ & $1.87~(0.00)$ & $0.94~(0.00)$ \\
%		& SEED & $16.62~(0.44)$ & $16.86~(0.35)$  & $1.23~(0.13)$  \\
%        & RCGL & $18.65~(0.74)$ & $15.97~(0.67)$  & $0.00~(0.00)$\\
%		& COSS & $7.51~(0.15)$ & $7.92~(0.15)$   & $0.00~(0.00)$  \\
%		\hline
%		\multirow{3}{*}{$400$}
%		%&SESS$^*$ & $1.87~(0.00)$ & $0.76~(0.00)$ \\
%		& SEED & $19.09~(1.26)$ & $17.73~(0.85)$  & $1.56~(0.17)$ \\
%        & RCGL & $20.97~(0.58)$ & $20.50~(0.85)$  & $0.00~(0.00)$\\
%		& COSS & $8.05~(0.44)$ & $8.24~(0.38)$  & $0.00~(0.00)$  \\
%		\hline
%		\multirow{3}{*}{$600$}
%		%&SESS$^*$ & $1.87~(0.00)$ & $0.78~(0.00)$ \\
%		& SEED & $27.63~(2.45)$ & $23.12~(21.68)$  & $3.21~(0.20)$ \\
%        & RCGL & $19.94~(0.49)$ & $19.50~(0.44)$  & $0.00~(0.00)$ \\
%		& COSS & $7.65~(0.35)$ & $7.78~(0.28)$  & $0.00~(0.00)$  \\
%		\hline
%        \multirow{3}{*}{$800$}
%		%&SESS$^*$ & $1.88~(0.00)$ & $0.75~(0.00)$ \\
%		& SEED & $28.66~(2.21)$ & $23.45~(1.95)$  & $3.68~(0.31)$ \\
%        & RCGL & $19.59~(0.33)$ & $20.53~(0.27)$  & $0.00~(0.00)$\\
%        & COSS & $9.44~(0.03)$ &$9.51~(0.03)$   & $0.00~(0.00)$ \\
%        \hline

	\end{tabular}
	\label{tab:synth1}
\end{table}

To compare the aforementioned methods, we consider the same performance measures as suggested in \cite{bahadori2016scalable}. The first two measures are the Normalized Estimation Error (NEE) and the Rank recovery Error (RE), defined as $\rm NEE(\widehat{\bC}) = \|\widehat{\bC} - \bC^*\|_F / \|\bC^*\|_F$ and $\rm RE(\widehat{\bC}) = |\mathrm{rank}(\widehat{\bC}) - \mathrm{rank}(\bC^*)|$, respectively. Based on an independent test sample of size 10000, the third measure is Normalized Prediction Error (NPE) defined as $\rm NPE(\widehat{\bC}) = \|\bY_{\text{test}} - \bX_{\text{test}} \widehat{\bC}\|_F / \|\bY_{\text{test}}\|_F$. Table \ref{tab:synth1} summarizes the simulation results of these three performance measures for additive, multiplicative error and missing data cases. In view of NPE and NEE in Tables \ref{tab:synth1}, it is clear that the performance of COSS is among the best in terms of either prediction or estimation accuracy.

\section{Discussion}\label{sec5}
In this paper, we have introduced a new methodology COSS to achieve scalable and interpretable estimation for multi-response error-in-variables regression under both additive and multiplicative measurement errors. It takes full advantage of recently developed sequential sparse factor estimation and the nearest positive semi-definite matrix projection, thus enjoying stepwise convexity and scalability in large-scale association analyses. Both the established theoretical properties and numerical performances demonstrate that the proposed method enjoys nice estimation, prediction, rank recovery accuracy and high scalability under both additive and multiplicative measurement errors. Similar theoretical conclusions can also be extend to random sub-Gaussian designs similar as that in \cite{Li2020}. It would be of interest to study several extensions of COSS to more general model settings such as the time series model and the generalized linear model, which are beyond the scope of the current paper and demands future studies.

\section*{Appendix}
\noindent \textbf{Proof of Theorem \ref{theo1}.}
%\subsection{Proof of Theorem ~\ref{theo1}}%main results
Before the proof of Theorem \ref{theo1}, we need to introduce a few additional notations used subsequently in the proofs that $\bleta_k = \widehat{\bZ}_k - \bX \bu^*_k$, $B = \max_{1 \leq i \leq r^*} \|\bu^*_i \|_{\infty}$. Denote $S_k = \{1,\dots,s_k\}$ the true support set of $k$-th singular vectors $\bu^*_k$ and write $\bu^*_k = (\bu^{*\top}_{S_k}, \bzero^\top)^\top$ and $\bX = (\bX_{S_k}, \bX_{S_k^c})$. Then we can rewrite $\bX \bu^*_k$ as $\bX_{S_k} \bu^*_{S_k}$ with the components of $\bu^*_{S_k}$ all non-zero. Since the following argument applies to vectors and unit rank matrix for any $k$, $1 \leq k \leq r^*$, we drop the index $k$ for ease of notational presentation. The proof for the bounds on the four quantities is composed of three parts.
%\textit{Step 1. Deriving the uniform bounds on $\|\widehat{\bu} - \bu^*\|_2$ and $\|\bX \widehat{\bu} - \bX \bu^*\|_2/\sqrt{n}$}.

Part 1: Deriving the uniform bounds on $\|\widehat{\bu} - \bu^*\|_2$ and $\|\bX \widehat{\bu} - \bX \bu^*\|_2/\sqrt{n}$.
%\noindent \textbf{Part 1: Deriving the uniform bounds on $\|\widehat{\bu} - \bu^*\|_2$ and $\|\bX \widehat{\bu} - \bX \bu^*\|_2/\sqrt{n}$.}
To ease readability, we will finish the proof in four steps.

Step 1. Denote $\widehat{\bdelta} = \widehat{\bu} - \bu^*$ the estimation error. Since the objective function (\ref{goal}) is convex, the global optimality of $\widehat{\bu}$ implies
\begin{align*}
\frac{1}{2}\widehat{\bu}^\top \widetilde{\bSig} \widehat{\bu} - \widetilde{\brho}^\top \widehat{\bu} + \widetilde{\lambda} \|\widehat{\bu}\|_1 \le \frac{1}{2}{\bu^*}^\top \widetilde{\bSig} \bu^* - \widetilde{\brho}^\top \bu^* + \widetilde{\lambda} \|\bu^*\|_1.
\end{align*}
By some simple calculation, we see that
\begin{align}\label{delta}
\frac{1}{2} \widehat{\bdelta}^\top \widetilde{\bSig} \widehat{\bdelta} + \widetilde{\lambda} \|\widehat{\bu}\|_1 \le \| \widehat{\bdelta} \|_1 \| \widetilde{\brho} - \widetilde{\bSig} \bu^* \|_\infty + \widetilde{\lambda} \| \bu^* \|_1.
\end{align}
By the triangular inequality, we have
\begin{align*}
\| \widetilde{\brho} - \widetilde{\bSig} \bu^* \|_\infty \le \| \widetilde{\brho} - \brho \|_\infty + \|{\brho} - \bSig \bu^* \|_\infty + \| (\widetilde{\bSig} - \bSig) \bu^* \|_\infty.
\end{align*}

We will then bound the three terms on the right hand successively. For the first term, by the union bounds (\ref{con:rho}) on $\widetilde{\brho}$ in Proposition \ref{rho}, it holds that for any $\widetilde{\lambda} > 0$,
\begin{align*}
\Pr (\| \widetilde{\brho} - \brho \|_\infty > \frac{\widetilde{\lambda}}{6}) & \le p \Pr (| \widetilde{\brho}_j - \brho_j | > \frac{\widetilde{\lambda}}{6}) \le p C \exp(-\frac{c n \widetilde{\lambda}^2 }{s^{2} \zeta}) .
\end{align*}
To bound the second term, by the the union bounds (\ref{con:sig}) on $\widetilde{\bSig}$, we have
\begin{align*}
\Pr (\| (\widetilde{\bSig} - \bSig) \bu^* \|_\infty > \frac{\widetilde{\lambda}}{6}) \le  p^2 \Pr ( sB |\widehat{\bSig}_{ij} - \bSig_{ij} | > \frac{\widetilde{\lambda}}{12})  \le p^2 C \exp(- \frac{cn\widetilde{\lambda}^2}{s^2 B^2 \zeta}).
 \end{align*}
For the last term, by the definition of $\brho$, $\bSig$ and notation $\bleta = \widehat{\bZ} - \bX \bu^*$, we get ${\brho}-\bSig \bu^* = \bX^\top \bleta/n$. Note that $\|\bx_j\|_2 /\sqrt{n} = 1$. Hence it follows from Lemma \ref{lambda} that for any $\widetilde{\lambda}/6 > 4\sqrt{\lambda_1}\gamma_u(\sqrt{n}+\sqrt{q})/(d_\lambda\sqrt{nq})$,
\begin{align*}
\Pr (\|{\brho}-\bSig \bu^* \|_\infty > \frac{\widetilde{\lambda}}{6}) & = \Pr (\| \frac{1}{n} \bX^\top \bleta \|_\infty > \frac{\widetilde{\lambda}}{6}) \le p \Pr (\frac{1}{\sqrt{n}}\|\bleta\|_2> \frac{\widetilde{\lambda}}{6}) \le p \exp(-\frac{n}{2}).
\end{align*}

Combing the three terms and redefining $\zeta = \max(\zeta,B^2\zeta)$, we have $\Pr (\|\widetilde{\brho} - \widetilde{\bSig} \bu^* \|_\infty )\le 1 - C p^2\exp(- cn\widetilde{\lambda}^2/s^{2}\zeta)$. For simplicity, we introduce the event $ \mathcal{F} = \{\|\widetilde{\brho} - \widetilde{\bSig} \bu^* \|_\infty \le \widetilde{\lambda}/2\}$ and assume that all our discussion will be conditioning on this new event $\mathcal{F}$ hereafter. Thus the remainder proofs hold simultaneously with probability at least $\Pr (\mathcal{F}) = 1 - C p^2\exp(- cn\widetilde{\lambda}^2/ s^{2}\zeta)$. Consequently, plugging $\|\widetilde{\brho} - \widetilde{\bSig} \bu^* \|_\infty \le \widetilde{\lambda} / 2$ into (\ref{delta}) yields
\begin{align}\label{goal11}
\frac{1}{2}\widehat{\bdelta}^\top \widetilde{\bSig} \widehat{\bdelta} + \widetilde{\lambda} \|\widehat{\bu}\|_1 \le \frac{\widetilde{\lambda}}{2}\| \widehat{\bdelta} \|_1 + \widetilde{\lambda}\| \bu^*\|_1.
\end{align}
%-----------------------------------------------------------------------------------------------------------------
Step 2. To continue, we need to control the term $\| \bu^* \|_1$ in (\ref{goal11}). Note that $\widehat{\bdelta}_{S^c} = \widehat{\bu}_{S^c}$ and $\|\bu^*\|_1 = \|\bu^*_{S}\|_1$. Therefore, (\ref{goal11}) together with the inequality $\|x\|_1 = \|x_S\|_1 + \| x_{S^c}\|_1$ for any vector $x$ and equality $\bu^*_{S_c} = 0$ entails that
\begin{align*}
\frac{1}{2} \widehat{\bdelta}^\top \widetilde{\bSig} \widehat{\bdelta} + \widetilde{\lambda} \|\widehat{\bu}_S \|_1 + \widetilde{\lambda} \|\widehat{\bdelta}_{S^c} \|_1 \le \frac{\widetilde{\lambda}}{2} \|\widehat{\bdelta}_{S} \|_1 + \frac{\widetilde{\lambda}}{2} \|\widehat{\bdelta}_{S^c} \|_1 + \widetilde{\lambda} \|\bu^*_S \|_1.
\end{align*}
Meanwhile, since $\|\widehat{\bu}_S \|_1 \geq \|\bu^*_S \|_1 - \|\widehat{\bdelta}_S \|_1$, it holds that
\begin{align}\label{goal2}
\widehat{\bdelta}^\top \widetilde{\bSig} \widehat{\bdelta} + \widetilde{\lambda} \|\widehat{\bdelta}_{S^c} \|_1 \leq 3 \widetilde{\lambda} \| \widehat{\bdelta}_S\|_1.
\end{align}
%Denote $S_k = (1,2,\dots,s_k)$ the true support set of $k$-th singular vectors $\bu^*_k$ and write $\bu^*_k = (\bu^{*T}_{S_k}, \bzero^T)^T$ and $\bX = (\bX_{S_k}, \bX_{S_k^c})$. Then we can rewrite $\bX \bu^*_k$ as $\bX_{S_k} \bu^*_{S_k}$ with the components of $\bu^*_{S_k}$ all non-zero.
Through the above two steps, we obtain the inequality (\ref{goal2}). Now we are ready to show the following analysis.

Step 3. %We now proceed to bound the $\widehat{\bdelta}^T \widetilde{\bSig} \widehat{\bdelta}$.
Denote by $D = \bSig - \widetilde{\bSig}$. It follows from the inequality $\|\bu_S \|_1 \leq \sqrt{s}\|\bu_S \|_2$ and the inequality (\ref{goal2}) that
\begin{align}\nonumber
& \widehat{\bdelta}^\top  \bSig \widehat{\bdelta} +  \widetilde{\lambda} \|\widehat{\bdelta} \|_1 = \widehat{\bdelta}^\top \widetilde{\bSig} \widehat{\bdelta} + \widetilde{\lambda} \|\widehat{\bdelta}_S \|_1 + \widetilde{\lambda} \|\widehat{\bdelta}_{S^c} \|_1 + \widehat{\bdelta}^\top  D \widehat{\bdelta} \\\label{step3-1}
& \leq 4 \widetilde{\lambda} \| \widehat{\bdelta}_S\|_1 + \widehat{\bdelta}^\top  D \widehat{\bdelta} \leq 4 \widetilde{\lambda} \sqrt{s} \| \widehat{\bdelta}_S\|_2 + \widehat{\bdelta}^\top  D \widehat{\bdelta} \leq 4 \widetilde{\lambda} \sqrt{s} \| \widehat{\bdelta}\|_2 + \widehat{\bdelta}^\top  D \widehat{\bdelta}.
\end{align}
Combining the Restricted Eigenvalue Condition \ref{cond5} with $\|\widehat{\bdelta}_{S^c} \|_1 \leq 3\|\widehat{\bdelta}_S \|_1$ and the inequality $4ab \leq a^2/4 + 16 b^2$ yields
\begin{align}\label{step3-2}
4 \widetilde{\lambda} \sqrt{s} \| \widehat{\bdelta}\|_2 + \widehat{\bdelta}^\top  D \widehat{\bdelta} \leq \frac{\widehat{\bdelta}^\top  \bSig \widehat{\bdelta} }{4} + \frac{16\widetilde{\lambda}^2 s}{\Omega} + |\widehat{\bdelta}^\top  D \widehat{\bdelta}|.
\end{align}
By similar arguments, the last term $|\widehat{\bdelta}^\top  D \widehat{\bdelta}|$ is bounded as follows,
\begin{align}\label{step3-3}
|\widehat{\bdelta}^\top  D \widehat{\bdelta}| \leq \|D \|_{\max}( \|\widehat{\bdelta}_S \|_1 + \|\widehat{\bdelta}_{S^c} \|_1)^2 \leq 16s \| D\|_{\max} \|\widehat{\bdelta} \|_2^2.
\end{align}
The inequality (\ref{sig2222}) in Lemma \ref{sig_inequality} together with the inequality (\ref{con:sig}) in Proposition \ref{rho} gives that for any $\epsilon \leq \min(\epsilon_0, \Omega/ 64s)$
\begin{align*}
\Pr (16s \| D \|_{\max} \geq \frac{\Omega}{4}) \leq p^2 \max_{ij} \Pr ( |\widehat{\bSig}_{ij} - \bSig_{ij} | \geq \frac{\Omega}{64s}) \leq p^2 \exp( - cn \Omega^2 / s^2 \zeta).
\end{align*}

We now plug the inequality (\ref{step3-3}) and (\ref{step3-2}) into (\ref{step3-1}). Then it holds that with probability at least $ 1 - p^2 \exp( - cn \Omega^2 / s^2 \zeta) - p^2 \exp( - cn\widetilde{\lambda}^2 / s^2 \zeta)$
\begin{align*}
\widehat{\bdelta}^\top  \bSig \widehat{\bdelta} +  \widetilde{\lambda} \|\widehat{\bdelta} \|_1 \leq \frac{\widehat{\bdelta}^\top  \bSig \widehat{\bdelta}}{4} + \frac{16\widetilde{\lambda}^2 s}{\Omega} + \frac{\Omega}{4} \|\widehat{\bdelta} \|_2.
\end{align*}
Applying the Restrict Eigenvalue Condition \ref{cond5} again yields $ \Omega/2\| \widehat{\bdelta}\|_2^2 + \widetilde{\lambda}\| \widehat{\bdelta}\|_1 \le 16\widetilde{\lambda}^2s/\Omega$. Therefore, for any $\epsilon < \min(\epsilon_0,\Omega/64s)$, with probability at least $ 1 - p^2 \exp( - cn \Omega^2 / s^2 \zeta) - p^2 \exp( - cn
\widetilde{\lambda}^2 / s^2 \zeta)$, we have
\begin{align}
\nonumber
&\| \widehat{\bdelta}\|_1 = \|\widehat{\bu} - \bu^*\|_1 \le \frac{16 s \widetilde{\lambda}}{\Omega}\\\label{u}
&\| \widehat{\bdelta}\|_2 = \|\widehat{\bu} - \bu^*\|_2 \le \frac{4 \sqrt{2 s } \widetilde{\lambda}}{\Omega}. % = C_{u}\sqrt{s} \Big(\frac{\sqrt{n} + \sqrt{q}}{\sqrt{nq}}\Big), \ C_{u} = \frac{4 \sqrt{2} \widetilde{C}}{\Omega}.
\end{align}

Step 4. An application of the triangular inequality and the inequality (23) in \cite{sun2012scaled}, which is derived from the Karush $-$ Kuhn $-$ Tucker condition, yields
\begin{align*}%\label{xtz}
\frac{2}{n} \|\bX \widehat{\bu} - \bX \bu^*\|^2_2 \leq 2 \widetilde{\lambda} (\|\bu^*\|_1 - \|\widehat{\bu}\|_1 ) + \frac{2}{n}\|\bX^\top (\widehat{\bZ} - \bX \bu^*)\|_\infty \cdot \| \bu^* - \widehat{\bu}\|_1.
\end{align*}
%\begin{align*}
% & \frac{2}{n} \|\bX\widehat{\bu} - \bX\bu^*\|^2_2 \leq 2 \widetilde{\lambda} (\|\bu^*\|_1 - \|\widehat{\bu}\|_1 ) + \frac{2}{n}\|\bX^T (\widehat{\bZ} - \bX \bu^*)\|_\infty \cdot \| \bu^* - \widehat{\bu}\|_1 \\
%                              & \leq 2 \widetilde{\lambda} \| \bu^* - \widehat{\bu}\|_1 \left( 1 + \frac{\xi - 1}{\xi+1} \right) = \frac{4 \widetilde{\lambda} \xi}{\xi+1} \| \widehat{\bu} - \bu^*\|_1 \leq \frac{8 \widetilde{\lambda}^2 \xi^2 s}{(\xi + 1)^2 F_1(\xi,S)}.
%\end{align*}
Note that $\|\bX_{j}\|_{2} = \sqrt{n}$. When $\widetilde{\lambda} = \max\{s \sqrt{(\zeta \log p)/n}$, $ \widetilde{C} (\sqrt{n} + \sqrt{q})/ \sqrt{nq} \}$ with $\widetilde{C} > 24\sqrt{\lambda_1}\gamma_u/d_{\lambda}$, it follows from Lemma \ref{lambda} that with probability at least $ 1 - \exp(\frac{n}{2})$,
\begin{align*} %\label{XTZ}
   \frac{1}{n} \|\bX^{\top}(\widehat{\bZ} - \bX\bu^*)\|_{\infty} \leq \max\limits_{1 \leq j\leq p} \frac{\|\bX_{j}\|_{2}}{\sqrt{n}} \cdot \frac{\|\widehat{\bZ} - \bX\bu^*\|_{2}}{\sqrt{n}} < \widetilde{\lambda}.
\end{align*}
By combining the above two inequalities, we obtain the desired error bound
\begin{align*}
\frac{1}{\sqrt{n}} \|\bX\widehat{\bu} - \bX\bu^*\|_2 \leq \frac{4 \sqrt{2 s}}{\sqrt{\Omega}} \widetilde{\lambda},
\end{align*}
%where $\widetilde{C}_u = 4 \sqrt{2}\widetilde{C}/\sqrt{\Omega}$.
It completes the first part of the proof.

%\textit{Step 2. Deriving the uniform bound on $\|\widehat{\bv} - \bv^*\|_2/\sqrt{q}$}.
Part 2: Deriving the uniform bound on $\|\widehat{\bv} - \bv^*\|_2/\sqrt{q}$. From the definition of $\bv^* = (1/n) \bY^{*\top} \bX \bu^*$, $\widehat{\bv} = (1/n) \bY^\top \widehat{\bZ}$ and assumed notation $\bleta = \widehat{\bZ} - \bX \bu^*$, we have
%With estimated eigenvalue vector $\widehat{\bZ}$ and noise version of $\bY^*$, the corresponding right singular vectors $\widehat{\bv}$ can be estimated as
%\begin{equation*}
%\widehat{\bv} = \frac{1}{n}\bY^T \widehat{\bZ}.
%\end{equation*}
%it gives
\begin{align*}
\|\widehat{\bv} - \bv^{*}\|_2  = \frac{1}{n} \|{\bY^*}^\top \bleta  - \bE^{\top}\widehat{\bZ} \|_2 \leq \frac{\|\bY^*\|_2 \cdot \|\bleta\|_2}{n} + \frac{\|\bE\|_2 \cdot \| \widehat{\bZ} \|_2}{n}.
\end{align*}
It is easy to observe that $\|\bY^*\|_2 = \sqrt{nq \lambda_1}$ since $nq \lambda_k = \|\bX \bC^*_k\|_F^2$. Thus, combing the equality $\|\widehat{\bZ} \|_2 = \sqrt{n}$ and the upper bound (\ref{E_1}) for $\|\bE\|_2$ in Lemma \ref{E_bound}, (\ref{etaa}) for $\|\bleta \|_2$ in Lemma \ref{lambda}, with probability at least $1 - 2\exp(-\frac{n}{2})$ we can get
\begin{align*}%\label{v}
\frac{1}{\sqrt{q}} \|\widehat{\bv} - \bv^{*}\|_2 & < \sqrt{\lambda_1} \cdot \widetilde{C} \Big(\frac{\sqrt{n} + \sqrt{q}}{\sqrt{nq}}\Big) + 2\gamma_u \Big(\frac{\sqrt{n} + \sqrt{q}}{\sqrt{nq}}\Big) = C_v \Big(\frac{\sqrt{n} + \sqrt{q}}{\sqrt{nq}}\Big),
\end{align*}
where $ C_v = (\sqrt{\lambda_1} \cdot \widetilde{C}+ 2\gamma_u)$.
%\begin{align}\label{v bound}
%\frac{1}{\sqrt{q}} \|\widehat{\bv} - \bv^{*}\|_2 < C_v \Big(\frac{\sqrt{n} + \sqrt{q}}{\sqrt{nq}}\Big),
%\end{align}
%where $C_v = \sqrt{\lambda_1} \cdot \widetilde{C}+ 2\gamma_u$.
%\textit{Step 3. Deriving the uniform bounds on $\|\widehat{\bC} - \bC^*\|_2/\sqrt{q}$ and $\|\bX\widehat{\bC} - \bX \bC^*\|_2/\sqrt{nq}$}.

Part 3: Deriving the uniform bounds on $\|\widehat{\bC} - \bC^*\|_2/\sqrt{q}$ and $\|\bX\widehat{\bC} - \bX \bC^*\|_2/\sqrt{nq}$. By definitions of the unit rank matrices $\bC^*$ and $\widehat{\bC}$, we have
\[\bC^* - \widehat{\bC} = \bu^* \bv^{*\top} - \widehat{\bu} \widehat{\bv}^\top = (\bu^* - \widehat{\bu})\bv^{*\top} + \widehat{\bu}(\bv^{*} - \widehat{\bv})^\top.\]
Therefore, Condition \ref{conduv} together with the triangular inequality $\|\widehat{\bu}\|_2 \leq \|\bu^*\|_2 + \|\widehat{\bu} - \bu^*\|_2$, the upper bound (\ref{u}) for $\widehat{\bu}$ and the above upper bound for $\widehat{\bv}$ entails that for sufficiently large $n$,
\begin{align*}
\frac{1}{\sqrt{q}}\|\widehat{\bC} - \bC^*\|_F < \frac{1}{\sqrt{q}} \|\bu^* - \widehat{\bu}\|_2 \cdot \|\bv^{*}\|_2 + \frac{1}{\sqrt{q}}\|\widehat{\bu}\|_2 \cdot \|\bv^{*} - \widehat{\bv}\|_2 < (V C_{u} + U C_{v}) \sqrt{s} \widetilde{\lambda},
\end{align*}
where $C_{u} = 4\sqrt{2}/\Omega$. For the prediction error bound of the unit rank matrix, applying similar analysis, with sufficiently large $n$, we have
\begin{align}
 \frac{1}{\sqrt{nq}} \|\bX(\bC^* - \widehat{\bC})\|_F < \frac{1}{\sqrt{nq}} \|\bX(\bu^* - \widehat{\bu})\|_2 \cdot \|\bv^{*}\|_2 + \frac{1}{\sqrt{nq}}\|\bX \widehat{\bu}\|_2 \cdot \|\bv^{*} - \widehat{\bv}\|_2 < (V \widetilde{C}_{u} + C_{v}) \sqrt{s} \widetilde{\lambda}. \label{XC}
\end{align}

\medskip

\noindent \textbf{Proof of Theorem \ref{theo2}.}
%\subsection{Proof of Theorem ~\ref{theo2}}
The proof is composed of three steps. We first derive the deterministic conclusion that $\mathcal{L}(k - 1) - \mathcal{L}(k) = \widehat{\lambda}_k$ for any $k \geq 0$, then verify that information criterion (\ref{rank}) will keep decreasing until the estimated rank reaches the true rank $r^*$ and start increasing with high probability in the second and third step respectively.
%We will show that $\mathcal{L}(k - 1) - \mathcal{L}(k) = \widehat{\lambda}_k$.

Step 1. Note that $\widehat{\bZ}_k$ is the $k$-th eigenvector of the problem (\ref{step1}) corresponding to the $k$-th eigenvalue $\widehat{\lambda}_k$. By the definition, we have $\widehat{\lambda}_k = \widehat{\bZ}_k^\top \bY \bY^\top \widehat{\bZ}_k/n^2 q$.
%It satisfy $\frac{1}{nq} \bY\bY^T \widehat{\bZ}_k = \widehat{\lambda}_k \widehat{\bZ}_k$. By some simple calculation, we have
%\begin{align*}
%\widehat{\lambda}_k = \frac{\widehat{\bZ}_k^T \bY \bY^T \widehat{\bZ}_k}{nq \widehat{\bZ}_k^T \widehat{\bZ}_k} = \frac{\widehat{\bZ}_k^T \bY \bY^T \widehat{\bZ}_k}{n^2 q}.
%\end{align*}
Since $\mathcal{L}(k) = \frac{1}{nq}\|\bY - \widehat{\bY}_{k}\|_F^2$ with $\widehat{\bY}_{k} = \sum_{j=1}^{k} \widehat{\bZ}_j \widehat{\bv}_j^\top$. In the $k$th step, applying the orthogonality between different eigenvectors $\widehat{\bZ}_k$ and replacing $\widehat{\bv}_k$ with $\widehat{\bv}_k = \bY^\top \widehat{\bZ}_k/n$ yield
\begin{align*}
\mathcal{L}(k - 1) - \mathcal{L}(k) = \frac{1}{nq}\left(2\left\langle \bY - \widehat{\bY}_{k - 1},  \widehat{\bZ}_k \widehat{\bv}_k^{\top}\right\rangle - \|\widehat{\bZ}_k \widehat{\bv}_k^{\top}\|_F^2 \right) = \widehat{\lambda}_k.
\end{align*}

We now analyze the information criterion (\ref{rank}). Some algebra gives
\begin{align}\label{tar}
\mathcal{C}({k - 1}) - \mathcal{C}({k}) = \sqrt{n} \log (\mathcal{L}(k - 1)/ \mathcal{L}(k)) - \log n.
\end{align}
Since $1 - 1/x \leq \log(x) \leq x-1 $ for $x >0$, the lower and upper bounds on $\log (\mathcal{L}(k - 1) / \mathcal{L}(k))$ can be provided by
\begin{align}\label{log}
\frac{\mathcal{L}(k - 1) - \mathcal{L}(k)}{\mathcal{L}(k - 1)} \leq \log \left( \frac{\mathcal{L}(k - 1)}{\mathcal{L}(k)}\right) \leq \frac{\mathcal{L}(k - 1) - \mathcal{L}(k)}{\mathcal{L}(k)}.
\end{align}
%Under these results, we are now ready to derive the following steps.

Step 2. We show that $\mathcal{C}({k - 1}) > \mathcal{C}({k})$ when $1 \leq k \leq r^*$. Under Condition \ref{cond2}-\ref{cond3}, using the perturbation bound in Lemma \ref{lambda} for the eigenvalues $\widehat{\lambda}$, with probability at least $1 - \exp(-n/2)$ we have
\begin{align}\label{boundL1}
\mathcal{L}(k - 1) - \mathcal{L}(k) = \widehat{\lambda}_k > \lambda_k - C \Big(\frac{\sqrt{n} + \sqrt{q}}{\sqrt{nq}}\Big).
\end{align}
By the definition of $\mathcal{L}(k - 1)$, it holds that
\begin{align*}
\sqrt{\mathcal{L}(k - 1)} \leq \frac{1}{\sqrt{nq}}\Big(\sum_{j=1}^{k-1}\|\bX \bu_j^* \bv_j^{*\top} - \widehat{\bZ}_j \widehat{\bv}_j^\top\|_F + \sum_{j=k}^{r^*}\|\bX \bu_j^* \bv_j^{*\top}\|_F + \|\bE\|_F \Big).
\end{align*}

We will then bound the above three terms respectively. For the first term, by the triangle inequality, we have
\begin{align*}
 \frac{1}{\sqrt{nq}} \|\bX \bu_j^* \bv_j^{*\top} - \widehat{\bZ}_j \widehat{\bv}_j^\top\|_F  \leq  \frac{1}{\sqrt{nq}} \|\bX \bu_j^* \bv_j^{*\top} - \widehat{\bZ}_j \bv_j^{*\top}\|_F +  \frac{1}{\sqrt{nq}} \|\widehat{\bZ}_j \bv^{*\top} - \widehat{\bZ}_j \widehat{\bv}_j^\top\|_F .
\end{align*}
By the inequality (\ref{etaa}) about error bound in Lemma \ref{lambda} and the assumed condition (\ref{conduv}), it holds that with probability at least $1 - \exp(-n/2)$ ,
\begin{align*}
& \frac{1}{\sqrt{nq}} \|\bX \bu_j^* \bv_j^{*\top} - \widehat{\bZ}_j \bv_j^{*\top}\|_F \leq \frac{4 V \sqrt{\lambda_1} \gamma_u} {d_\lambda} \cdot \frac{\sqrt{n} + \sqrt{q}}{\sqrt{nq}} \\
& \frac{1}{\sqrt{nq}} \|\widehat{\bZ}_j \bv^{*\top} - \widehat{\bZ}_j \widehat{\bv}_j^T\|_F \leq C_v \Big(\frac{\sqrt{n} + \sqrt{q}}{\sqrt{nq}}\Big).
\end{align*}

Combining above two inequalities leads to
\begin{align*}
\sum_{j=1}^{k-1} \frac{\|\bX \bu_j^* \bv_j^{*\top} - \widehat{\bZ}_j \widehat{\bv}_j^\top\|_F}{\sqrt{nq}} < (k - 1) C_z \Big(\frac{\sqrt{n} + \sqrt{q}}{\sqrt{nq}}\Big),  \ C_z = 4 V \sqrt{\lambda_1} \gamma_u / d_\lambda + C_v.
\end{align*}
For the last two terms, since $ \|\bX \bu_j^* \bv_j^{*\top}\|_F = \sqrt{nq \lambda_k}$, the inequality (\ref{E_2}) in Lemma \ref{E_bound} entails that with probability at least $1 - \exp(-n/2)$,
\begin{align*}
\frac{1}{\sqrt{nq}} \sum_{j=k}^{r^*}\|\bX \bu_j^* \bv_j^{*\top}\|_F = \sum_{j=k}^{r^*} \sqrt{\lambda_k}, \quad \frac{1}{\sqrt{nq}} \|\bE\|_F \leq \gamma_u \left(1 + \frac{1}{\sqrt{q}} \right).
\end{align*}
Assume that $c_{\gamma}= \gamma_u /\sqrt{\lambda}_{r^*}$ are finite constant. Therefore, it follows from the above three terms that with probability at least $1 - c_0 \exp(n/2)$,
\begin{align*}
\sqrt{\mathcal{L}(k - 1)} \leq (r^* - k + 1 + c_{\gamma}) \sqrt{\lambda}_k + (\gamma_u + r^* C_z) \Big(\frac{\sqrt{n} + \sqrt{q}}{\sqrt{nq}}\Big),
\end{align*}
where the last inequality hold due to $ c_{\gamma} \sqrt{\lambda}_{r^*} \leq c_{\gamma} \sqrt{\lambda}_k$ and $c_0$ is a finite constant.

The above inequality together with the inequality (\ref{boundL1}) entails that %the lower bound of (\ref{log})
\begin{align*}
\sqrt{\frac{\mathcal{L}(k - 1) - \mathcal{L}(k)}{\mathcal{L}(k - 1)}} > = \frac{1}{r^* - k + 1 + c_{\gamma}} + O\left(r^* \frac{\sqrt{n} + \sqrt{q}}{\sqrt{nq}} \ \right).
\end{align*}
It follows immediately from the assumptions ${r^*} (\log n/\sqrt{n})^{1/2} = o(1)$, which yields ${r^*} ((\sqrt{n} + \sqrt{q})/\sqrt{nq})^{1/2} = o(1)$, that for sufficiently large $n$, we have $\sqrt{(\mathcal{L}(k - 1) - \mathcal{L}(k))/\mathcal{L}(k - 1)} >  1/(r^* - k + 1 + c_{\gamma}).$ We now plug the above inequality and (\ref{log}) into (\ref{tar}). Then it holds that
\begin{align*}
& \mathcal{C}({k - 1}) - \mathcal{C}({k}) = \sqrt{n} \log (\frac{\mathcal{L}(k - 1)}{\mathcal{L}(k)}) - \log n > \frac{\sqrt{n}}{(r^* - k + 1 + c_{\gamma})^2} - \log n  > 0,
\end{align*}
which means that the information criterion $\mathcal{C}({k})$ will keep decreasing until our selected rank $k$ equal to the true rank $r^*$.
\smallskip

Step 3. We show that $\mathcal{C}({k - 1}) < \mathcal{C}({k})$ when $k > r^*$. since $\lambda_k = 0$ for any $k > r^*$, the inequality (\ref{lambda_k}) in Lemma \ref{lambda} entails that,
\begin{align}\label{sig}
\mathcal{L}(k - 1) - \mathcal{L}(k) = \widehat{\lambda}_k < C \Big(\frac{\sqrt{n} + \sqrt{q}}{\sqrt{nq}}\Big).
\end{align}
Meanwhile, by the definition of $\mathcal{L}(k)$ and the triangular inequality, we have
\begin{align*}%\nonumber
\sqrt{\mathcal{L}(k)} \geq \frac{1}{\sqrt{nq}}\Big(\|\bE\|_F - \sum_{j=1}^{r^*}\|\bX \bu_j^* \bv_j^{*\top} - \widehat{\bZ}_j \widehat{\bv}_j^\top\|_F - \sum_{j = r^* + 1}^{k} \|\widehat{\bZ}_j \widehat{\bv}_j^\top\|_F\Big).
\end{align*}

By similar arguments as in \textit{step 1}, under the assumptions $r^* \{(\sqrt{n} + \sqrt{q})/\sqrt{nq} \}^{1/2} = o(1)$ and $r \{(\sqrt{n} + \sqrt{q})/\sqrt{nq}\}^{1/2} = o(1)$, for sufficient large $n$, combining the above three terms and the inequality (\ref{sig}) yields
%\begin{align*}
%\sqrt{\frac{\mathcal{L}(k - 1) - \mathcal{L}(k)}{\mathcal{L}(k)}} & < \frac{ \sqrt{C} \left( \frac{\sqrt{n} + \sqrt{q}}{\sqrt{nq}}\right)^{1/2}}{\gamma_l - (C_z + 2\gamma_l) \left( r^* \frac{\sqrt{n} + \sqrt{q}}{\sqrt{nq}} \right) - (r - r^*) \sqrt{C} \left( \frac{\sqrt{n} + \sqrt{q}}{\sqrt{nq}} \right)^{1/2}}.
%\end{align*}
%Then by  we have
\begin{align*}%\label{kbr}
\sqrt{\frac{\mathcal{L}(k - 1) - \mathcal{L}(k)}{\mathcal{L}(k)}} & < \frac{\sqrt{C}}{\gamma_l} \left( \frac{\sqrt{n} + \sqrt{q}}{\sqrt{nq}} \right)^{1/2}.
\end{align*}
Therefore, applying the inequality (\ref{log}) gives
\begin{align*}
\mathcal{C}({k - 1}) - \mathcal{C}({k}) \le \sqrt{n} \frac{\mathcal{L}(k - 1) - \mathcal{L}(k)}{\mathcal{L}(k)} -  \log n < \frac{C}{\gamma_l^2} \left( \frac{\sqrt{n} + \sqrt{q}}{\sqrt{q}} \right) - \log n < 0,
\end{align*}
where the last inequality is immediate from the assumption that $\sqrt{n}/(\sqrt{q} \log n) = o (1)$. It indicate that the information criterion will keep increasing once the estimated rank $k$ is larger than $r^*$. Consequently, with probability at least $1 - c_0 \exp(-n/2)$ for sufficiently large $n$, $\mathcal{C}({k})$ will attain its minimum value when $k = r^*$.
%This result means that our Integrating the probabilities above, Theorem \ref{theo2} hold with probability at least $1 - c_0 \exp(-n/2)$ with $c_0$ finite constant.
Thus, we finish the proof of Theorem \ref{theo2}.

\medskip

\noindent \textbf{Proof of Proposition \ref{rho}.}
%\subsection{Proof of Proposition ~\ref{rho}}
%we need to address two main issues before claiming the success of SESS
%Before showing the proofs for the results, some preparations are needed. First, we introduce a useful definition given in .
\medskip

\noindent \textbf{Definition.} (Sub-Gaussian random vectors) A random vector $\bw$ is said to be sub-Gaussian if there exists $\tau > 0$ such that
\begin{align}\label{sub-gaussion1}
\Pr (|\bv^\top (\bw - {\rm E}(\bw))| > t) \leq 2 \exp(-\frac{2t^2}{2 \tau^2})
\end{align}
for all $t > 0$ and $\|\bv\|_2 = 1$.

Second, we impose several regularity conditions for the multiplicative setup:
\begin{equation}\label{mul-cond}
\mathop{\max}\limits_{i,j}\lvert \bx_{ij}\rvert = X_{\max} < \infty, \quad \min\bmu_M = \mu_{\min} > 0,
\end{equation}
and introduce a few additional notations used subsequently in the proofs that $\bleta_k = \widehat{\bZ}_k - \bX \bu^*_k$, $B = \max_{1 \leq i \leq r^*} \|\bu^*_i \|_{\infty}$,
%\begin{align*}
%\bleta_k = \widehat{\bZ}_k - \bX \bu^*_k \quad, \quad  B = \max_{1 \leq i \leq r^*} \|\bu^*_i \|_{\infty},
%\end{align*}
$\ba_j$ denotes the $j$-th column of matrix $\bA$, $\mu_j$ denotes the $j$-th element of vector $\bmu$. Similar notations hold for $\bx_j$, $\bm_j$, $u_k^*$. Since the following argument applies to $\widetilde{\brho}_{{\rm add}}^k$, $\widetilde{\brho}_{{\rm mul}}^k$, $\widehat{\bZ}_k$, $\bu_k^*$, $\bleta_k$, $s_k$ and $S_k$ with any fixed $k$, $1\le k \le r^*$, we drop the index $k$ for notation clarity.

%Denote $S_k = (1,2,\dots,s_k)$ the true support set of $k$-th singular vectors $\bu^*_k$ and write $\bu^*_k = (\bu^{*T}_{S_k}, \bzero^T)^T$ and $\bX = (\bX_{S_k}, \bX_{S_k^c})$. Then we can rewrite $\bX \bu^*_k$ as $\bX_{S_k} \bu^*_{S_k}$ with the components of $\bu^*_{S_k}$ all non-zero.

Observe that the proof for the inequality (\ref{con:sig}) are exactly the same as that in \cite{datta2017}, so we present the proof for the inequality (\ref{con:rho}) here. The proof is composed of two parts.

Part 1: Proof for additive measurement error case. Note that it follows from the definition of vectors $\widetilde{\brho}_{{\rm add}} = 1/n \bW^\top \widehat{\bZ}$ that
\begin{align*}
\widetilde{\brho}_{{\rm add},j} - \brho_j = (\frac{1}{n}\bW^\top \widehat{\bZ})_j - (\frac{1}{n}\bX^\top \widehat{\bZ})_j  \frac{1}{n} \{\bA^\top (\bX_{S} \bu^*_{S} + \bleta)\}_j= \frac{1}{n}\ba_j^\top \bX_S \bu_S^*+\frac{1}{n} \ba_j^\top \bleta.
\end{align*}
We will bound the two terms on the right hand side successively.

For the first term, since $\|\bx_i\|_2 = \sqrt{n}$ and the entries of $\ba_j$ are independent and Sub-Gaussian with parameter at most $\tau^2$, applying inequality $|\ba_j^\top \bX_S \bu^*_S  | \leq \|\ba_j^\top \bX_S \|_1 \cdot \|\bu_S^*\|_{\infty}$ over all nodes j in the index sets yielding
\begin{align*}
\Pr (\lvert \frac{1}{n} \ba_j^\top \bX_S \bu_S^* \rvert > \frac{\epsilon}{2} ) \le \Pr (B \Sigma_{i=1}^s \lvert \frac{1}{n} \ba_j^\top \bx_i \rvert > \frac{\epsilon}{2} ) \le 2\exp(- \frac{n \epsilon^2}{8 s^2 \tau^2 B^2 }), %= Cexp(-cn\epsilon^2s^{-2} \zeta^{-1}),
\end{align*}
where the last inequality is immediate from Sub-Gaussian inequality (\ref{sub-gaussion1}).
%where the last inequality is an application of sub-Gaussian variable properties with sub-Gaussian parameter $\tau^2$, absolute constants $C$ and $c$, $\|\bx_i\|_2 = \sqrt{n}$. %and $\zeta = \tau^2 B^2$.

For the second term, since $|\ba_j^\top \bleta| \leq \| \ba_j\|_2 \cdot \| \bleta\|_2$ and $\|\ba_j\|_2 \leq \|\ba_j\|_1$, we have
\begin{align*}
\Pr (\lvert\frac{1}{n} \ba_j^\top \bleta \rvert > \frac{\epsilon}{2}) & \le \Pr(\frac{1}{\sqrt{n}}\lVert \ba_j \rVert_1 \cdot \frac{1}{\sqrt{n}} \lVert \bleta \rVert_2 > \frac{\epsilon}{2}) = \Pr (\frac{1}{\sqrt{n}} | \mathbf{1}_n^\top \ba_j | > \frac{\sqrt{n} \epsilon}{2\| \bleta \|_2}),
\end{align*}
where $\mathbf{1}_n$ denotes a vector whose elements are all equal to 1. By the Lemma \ref{lambda} in Section \ref{LEEMA}, we see that event $\|\bleta \|_2 \leq 4 \sqrt{\lambda_1} \gamma_u  (\sqrt{n} + \sqrt{q})/ d_\lambda \sqrt{q}$ holds with probability at least $1 - \exp(-n/2)$. Then together with inequality (\ref{prob}) in Lemma \ref{pro}, we have
\begin{align*}
\Pr (\frac{1}{\sqrt{n}} | \mathbf{1}_n^\top \ba_j |> \frac{\sqrt{n} \epsilon}{2\| \bleta \|_2}) \leq & \Pr ( \frac{1}{\sqrt{n}} | \mathbf{1}_n^\top \ba_j |  > \frac{\epsilon d_\lambda \sqrt{nq}}{8 \gamma_u \sqrt{\lambda_1} (\sqrt{n} + \sqrt{q})} ) \\
 & + \Pr ( \| \bleta\|_2 >  \frac{4 \sqrt{\lambda_1} \gamma_u  (\sqrt{n} + \sqrt{q})}{ d_\lambda \sqrt{q} } ).
\end{align*}

Hence it follows from the Sub-Gaussian inequality (\ref{sub-gaussion1}) that
\begin{align*}
\Pr (\lvert\frac{1}{n} \ba_j^\top \bleta \rvert > \frac{\epsilon}{2}) \le 2 \exp(-\frac{ q d_\lambda^2 n\epsilon^2}{128 \tau^2 \gamma_u^2 \lambda_1 (\sqrt{n} + \sqrt{q})^2} ) + \exp(-\frac{n}{2}).
\end{align*}
%
%\begin{align*}
%P(\frac{1}{\sqrt{n}} & | \mathbf{1}_n^T \ba_j |> \frac{\sqrt{n} \epsilon}{2\| \bleta \|_2}) \leq P( \frac{1}{\sqrt{n}} | \mathbf{1}_n^T \ba_j | > \frac{\epsilon d_\lambda \sqrt{nq}}{8 \gamma_u \sqrt{\lambda_1} (\sqrt{n} + \sqrt{q})} ) \\
%& + P( \| \bleta\|_2 >  \frac{4 \sqrt{\lambda_1} \gamma_u  (\sqrt{n} + \sqrt{q})}{ d_\lambda \sqrt{q} } ) \le 2 \exp(-\frac{\epsilon^2 d_\lambda^2 nq}{128 \tau^2 \gamma_u^2 \lambda_1 (\sqrt{n} + \sqrt{q})^2} ) + \exp(-\frac{n}{2}).
%\end{align*}
%By the assumption $q \geq n$,
By the assumption $\sqrt{n} + \sqrt{q} / \sqrt{nq} = o(1)$, combining the results of the above two terms with $\zeta = \max \{8 \tau^2 B^2, 128 \lambda_1 \gamma_u^{2} \tau^{-2} d_{\lambda}^{-2}\}$ and $C$ and $c$ generic positive constants, we have%$\widetilde{\brho}_{add}$ can satisfy (\ref{con:rho}),
\begin{align*}
\Pr ( | \widetilde{\brho}_{add,j} - \brho_j | > \epsilon) \leq C \exp (- c \epsilon^2 n s^{-2} \zeta^{-1}).
\end{align*}

%------------------------------------------------------------------------------------------------------------------------------------------------------------

%%%%%%%%%%%%%%%%%%%%%%%%%%%%%%%%%%%%%%%%%%%%%%%%%%%%%%%%%%%%%%%%%%%%%%%%%%%%%%%%
Part 2: Proof for multiplicative measurement error case. It follows from the definition of $\widetilde{\brho}_{{\rm mult}} = 1/n \bW^\top \widehat{\bZ} \oslash \bmu_M$ that
\begin{align*}
\widetilde{\brho}_{{\rm mult},j}-\brho_j  = (\frac{1}{n}\bW^\top \widehat{\bZ} \oslash \bmu_M)_j - (\frac{1}{n}\bX^\top \widehat{\bZ})_j = \frac{1}{n \mu_j} (\bw_j^\top - \mu_j \bx_j^\top)(\bX_{S} \bu^*_{S} + \bleta).
\end{align*}
%\begin{align*}
%\widetilde{\brho}_{mult,j}-\brho_j  & = (\frac{1}{n}\bW^T \widehat{\bZ} \oslash \bmu_M)_j - (\frac{1}{n}\bX^T \widehat{\bZ})_j = \frac{1}{n} (\bW_j - \bX_j)^T (\bX \bu^* + \bleta) / \mu_j\\
%                                    & = \frac{1}{n \mu_j} (\bw_j^T - \mu_j \bx_j^T)(\bX_{S} \bu^*_{S} + \bleta).
%\end{align*}
By the assumed condition $\min \bmu_M = \mu_{\min}$ in (\ref{mul-cond}) and $B = \max_{1 \leq i \leq r^*} \|\bu^*_i \|_{\infty}$, it holds that
\begin{equation}
 \begin{split}
 \nonumber
  | \widetilde{\brho}_{{\rm mult},j} - \brho_j| = & \frac{1}{\mu_j} \left[ \frac{1}{n}(\bm_j \odot \bx_j - \mu_j \bx_j)^\top ( \bSig_{k = 1}^{s} \bx_{k} u^*_{k} + \bleta ) \right] \\
              \le &\frac{B}{\mu_{\min}}\sum_{k=1}^s | \frac{1}{n}\sum_{i=1}^n \bx_{ij} \bx_{ik}(m_{ij} - \mu_j)|  + \frac{1}{\mu_{\min}} | \frac{1}{n}(\bw_j-\mu_j \bx_j)^\top \bleta |.
 \end{split}
\end{equation}
By similar arguments as in Part 1, we obtain the results as follows
\begin{align*}
 & \Pr (\frac{sB}{\mu_{\min}} | \frac{1}{n} \sum_{i=1}^n \bx_{ij} \bx_{ik}(m_{ij} - \mu_{j}) | \geq \frac{\epsilon}{2})\le 2 \exp(- \frac{\epsilon^2 \mu^2_{\min} n}{8 s^2 B^2 \tau^2 X_{\max}^4}). \\
 & \Pr (\frac{1}{\mu_{\min}} | \frac{1}{n}(\bw_j - \mu_j \bx_j)^\top \bleta | > \frac{\epsilon}{2}) \le 2 \exp(-\frac{ nq \epsilon^2 \mu^2_{\min}d_\lambda^2}{128\lambda_1 \gamma_u^2 \tau^2 (\sqrt{n}+\sqrt{q})^2 }) + \exp(\frac{n}{2}).
\end{align*}
Under the assumption $(\sqrt{n} + \sqrt{q}) / \sqrt{nq} = o(1)$, combining the two bounds gives
\begin{align*}
\Pr ( | \widetilde{\brho}_{{\rm mult},j}-\brho_j | > \epsilon) \leq C \exp (- c \epsilon^2 n s^{-2} \zeta^{-1}),
\end{align*}
where $\zeta = \max \{X_{\max}^4 \tau^2 B^2 \mu_{\min}^{-2} , \lambda_1 \gamma_u^2 \tau^{2} \mu_{\min}^{-2} d_\lambda^{-2}\}$ and $C$ and $c$ denote generic positive constants. %It conclude the proof for multiplicative error case. These two parts complete the proof of Proposition \ref{rho}.

\medskip

%%%%%%%%%%%%%%%%%%%%%%%%%%%%%%%%%%%%%%%%%%%%%%%%%%%%%%%%%%%%%%%%%%%%%%%%%%%%%%%%
%\subsection{Lemmas and their proofs} \label{LEEMA}
\medskip

\noindent \textbf{Lemmas and their proofs.} The following lemmas are used in the proof of the main theorems.

%The following lemma will be used in the proof of the main theorem,

%%%%%%%%%%%%%%%%%%%%%%%%%%%%%%%%%%%%%%%%%%%%%%%%%%%%%%%%%%%%%%%%%%%%%%%%%%%%%%%%

\begin{lemma}\label{lambda}
Consider the following eigenvalue problem and its perturbed variant:
\begin{align}\label{Y1}
\frac{1}{nq} \bY^* \bY^{*\top} \bZ = \lambda \bZ, \quad \frac{1}{nq} \bY \bY^{\top} \widehat{\bZ} = \widehat{\lambda} \widehat{\bZ},
\end{align}
where $\bX \bu^*_k$ and $\widehat{\bZ}_k$ are the $k$th eigenvectors of the problems (\ref{Y1}) with respect to the eigenvaluies $\lambda_k$ and $\widehat{\lambda}_k$, respectively. Assume that the conditions \ref{cond2}-\ref{cond3} hold and let $1 \le k \le r^*$. Then with probability at least $1 - \exp(-n/2)$, we have
\begin{align}\label{lambda_k}
| \widehat{\lambda}_k - \lambda_k | & \le C \frac{\sqrt{n} + \sqrt{q}}{\sqrt{nq}}, \\ \label{etaa}
\frac{1}{\sqrt{n}} \|\widehat{\bZ}_k - \bX \bu_k^*\|_2 & < \frac{4 \sqrt{\lambda_1} \gamma_u} {d_\lambda} ( \frac{\sqrt{n} + \sqrt{q}}{\sqrt{nq}}) .
\end{align}
where $C, \gamma_u$ and $d_{\lambda}$ are finite constant.
\end{lemma}
%\begin{proof}[\textit{Proof of Lemma~\ref{lambda}}]
\noindent {\bf Proof.} Since matrix $\bY$ is not affected by the measurement errors, the problems (\ref{Y1}) are exactly the same as those in \cite[Thereom 1]{Zheng19}. Thus based on the proof of \cite[Thereom 1]{Zheng19} we obtain the results in (\ref{lambda_k})-(\ref{etaa}). Please see \cite{Zheng19} for more discussion.
%\end{proof}

%%%%%%%%%%%%%%%%%%%%%%%%%%%%%%%%%%%%%%%%%%%%%%%%%%%%%%%%%%%%%%%%%%%%%%%%%%%%%%%%
\begin{lemma}\label{sig_inequality}
For any $\epsilon > 0$ we have
\begin{align}\label{sig2222}
\Pr (\|\widetilde{\bSig} - \bSig \|_{\max} \geq \epsilon) \leq p^2 \max_{i,j} \Pr (|\widehat{\bSig}_{i,j} - \bSig_{i,j} | \geq \frac{\epsilon}{2}).
\end{align}
\end{lemma}
%\begin{proof}[\textit{Proof of Lemma~\ref{sig_inequality}}]
\noindent {\bf Proof.} By the definition of $\widetilde{\bSig}$ in (\ref{semi-positive}), we have
\begin{align*}%\label{close}
\|\widetilde{\bSig} - \bSig \|_{\max} \leq \| \widetilde{\bSig} - \widehat{\bSig}\|_{\max} + \|\widehat{\bSig} - \bSig \|_{\max} \leq 2 \|\widehat{\bSig} - \bSig \|_{\max}.
\end{align*}
Further applying this inequality gives
\begin{align*}
\Pr (\|\widetilde{\bSig} - \bSig \|_{\max} \geq \epsilon) \leq \Pr (\|\widehat{\bSig} - \bSig \|_{\max} \geq \frac{\epsilon}{2}) \leq p^2 \max_{i,j} \Pr (|\widehat{\bSig}_{ij} - \bSig_{ij} | \geq \frac{\epsilon}{2}),
\end{align*}
which completes the proof of Lemma \ref{sig_inequality}.
%\end{proof}

%%%%%%%%%%%%%%%%%%%%%%%%%%%%%%%%%%%%%%%%%%%%%%%%%%%%%%%%%%%%%%%%%%%%%%%%%%%%%%%%
\begin{lemma}\label{pro}
Given two events A and B, we have
\begin{align}\label{prob}
\Pr (A) \le \Pr (A|B) + \Pr (B^c),
\end{align}
where $\Pr (A|B)$ is a conditional probability, $B^c$ denotes the opposite event of $B$.
\end{lemma}
%\begin{proof}[\textit{Proof of Lemma~\ref{pro}}]
\noindent {\bf Proof.} By simple probability calculations, we have
\begin{align*}
 \Pr (A) = \Pr (A\cap B) + \Pr (A\cap B^c) \le \Pr(A|B) \cdot \Pr (B) + \Pr (B^c) \le \Pr (A|B) + \Pr (B^c).
\end{align*}
Thus we finish the proof of Lemma \ref{pro}.
%\end{proof}
%%%%%%%%%%%%%%%%%%%%%%%%%%%%%%%%%%%%%%%%%%%%%%%%%%%%%%%%%%%%%%%%%%%%%%%%%%%%%%%%
\begin{lemma}\label{E_bound}
Under Condition \ref{cond3}, with probability at least $1- \exp(-n/2)$, the $n \times q$ random matrix $\bE = (\be_1, \ldots, \be_n )^\top$ with rows $\be_i$  i.i.d. $\sim  \mathcal{N} (\mathbf{0}, \mathbf{\Sigma}_E)$ satisfies
\begin{align} \label{E_1}
                                               &   \|\bE\|_{2}\leq\ \gamma_u (2\sqrt{n} + \sqrt{q}),  \\ \label{E_2}
 \gamma_l \left(1 - 2 / \sqrt{q}  \right) \leq &  \|\bE\|_F / \sqrt{nq} \leq \gamma_u \left(1 + 1 / \sqrt{q} \right).
\end{align}
\end{lemma}
%\begin{proof}[\textit{Proof of Lemma~\ref{E_bound}}]
\noindent {\bf Proof.} (i) The proof of the first inequality (\ref{E_1}). By simple calculation, we have ${\rm E}(\mathbf{\Sigma}_E^{-1/2} \be_i) = \bzero$ and ${\rm Cov}(\mathbf{\Sigma}_E^{-1/2} \be_i) = \bI$ with $\rm \bI$ identity matrix for any $i, 1\leq i \leq n$. Thus the entries of matrix $\mathbf{\Sigma}_E^{-1/2} \be_i$ are independent $\rm N (0,1)$  random variables.
%Thus, $ \mathbf{\Sigma}_E^{-1/2} \bveps_i$ are $q$-dimensional vectors with independent zero mean and unit variance entries for any $i, 1\leq i \leq n$.
%Combined with the standard random matrix theory \citep{rudelson2010non} that $\mathrm{E} \left(\| \mathbf{\Sigma}^{-1/2}\bE^T \|_2 \right) \leq \sqrt{n} + \sqrt{q}$,
An application of \cite[Lemma 3]{bunea2011optimal} yields
\begin{align*}
\Pr ( \| \mathbf{\Sigma}^{-1/2}\bE^\top \|_2 \geq \sqrt{n} + \sqrt{q} + t ) \leq \exp(-\frac{t^2}{2}),
\end{align*}
for any $t > 0$. Taking $t = \sqrt{n}$, the above inequality together with Condition \ref{cond2} gives
\begin{align*}
\|\bE\|_2 = \|\bE^\top\|_2 = \|\mathbf{\Sigma}^{1/2} \mathbf{\Sigma}^{-1/2} \bE^\top\|_2 \leq \|\mathbf{\Sigma}^{1/2}\|_2 \cdot \|\mathbf{\Sigma}^{-1/2}\bE^\top\|_2 \leq \gamma_u (2\sqrt{n} + \sqrt{q}),
\end{align*}
%\begin{small}
%\begin{align*}
%\|\bE\|_2 = \|\bE^T\|_2 = \|\mathbf{\Sigma}^{1/2} \mathbf{\Sigma}^{-1/2} \bE^T\|_2 \leq \|\mathbf{\Sigma}^{1/2}\|_2 \cdot \|\mathbf{\Sigma}^{-1/2}\bE^T\|_2 \leq \gamma_u (2\sqrt{n} + \sqrt{q}),
%\end{align*}
%\end{small}
where the last inequality holds with probability at least $1 - \exp(-\frac{n}{2})$.

(ii) The proof of the second inequality. Similarly, based on the event that $\mathbf{\Sigma}_E^{-1/2} \be_i$ is a $q \times n$ matrix with independent zero mean and unit variance entries, an application of the tail bound for $\chi^2$ distribution in \cite[Lemma 1]{laurent2000adaptive} gives
% with probability at least $1 - \exp(- n/2)$ we obtain
\begin{align}\label{E-upper}
\|\bE \mathbf{\Sigma}^{-1/2}\|_F^2/nq  \leq 1 + \sqrt{\frac{2}{q}} + \frac{1}{q} < \left(1 + \frac{1}{\sqrt{q}} \ \right)^2,
\end{align}
with probability at least $1 - \exp(- n/2)$. On the other hand, by Condition \ref{cond3} we can obtain that
\begin{align*}
\|\bE \mathbf{\Sigma}^{-1/2}\|_F^2 \geq \|\bE\|_F^2 \lambda_{\min}^2(\mathbf{\Sigma}^{-1/2}) \geq \|\bE\|_F^2/\gamma_{u}^2.
\end{align*}
The above inequality together with (\ref{E-upper}) gives $\|\bE\|_F/\sqrt{nq}  \leq \gamma_u \left(1 + 1/\sqrt{q} \right)$. Using the similar argument, it is easy to obtain the desired lower bound that $\|\bE\|_F / \sqrt{nq} \geq \gamma_l \left(1 - 2/\sqrt{q}  \right)$. Thus we finish the proof of Lemma~\ref{E_bound}.
%\end{proof}

%\section*{References}
\bibliography{mybibfile}

\begin{thebibliography}{29}
\expandafter\ifx\csname natexlab\endcsname\relax\def\natexlab#1{#1}\fi
\providecommand{\bibinfo}[2]{#2}
\ifx\xfnm\relax \def\xfnm[#1]{\unskip,\space#1}\fi
%Type = Article
\bibitem[{Baselmans et~al.(2019)Baselmans, Jansen and Ip}]{Baselmans2019}
\bibinfo{author}{B.~M.~L. Baselmans}, \bibinfo{author}{R.~Jansen},
  \bibinfo{author}{H.~F. Ip}, \bibinfo{title}{Multivariate genome-wide analyses
  of the well-being spectrum}, \bibinfo{journal}{Nature Genetics}
  \bibinfo{volume}{51} (\bibinfo{year}{2019}) \bibinfo{pages}{445--451}.
%Type = Article
\bibitem[{Belloni et~al.(2017)Belloni, Rosenbaum and Tsybakov}]{Belloni2017}
\bibinfo{author}{A.~Belloni}, \bibinfo{author}{M.~Rosenbaum},
  \bibinfo{author}{A.~B. Tsybakov}, \bibinfo{title}{Linear and conic
  programming estimators in high dimensional errors-in-variables models},
  \bibinfo{journal}{J. Roy. Statist. Soc. Ser. B} \bibinfo{volume}{79}
  (\bibinfo{year}{2017}) \bibinfo{pages}{939--956}.
%Type = Article
\bibitem[{Bickel and Ritov(1987)}]{Bickel1987}
\bibinfo{author}{P.~J. Bickel}, \bibinfo{author}{Y.~Ritov},
  \bibinfo{title}{Efficient estimation in the errors in variables model},
  \bibinfo{journal}{Ann. Statist.} \bibinfo{volume}{15} (\bibinfo{year}{1987})
  \bibinfo{pages}{513--540}.
%Type = Article
\bibitem[{Bickel et~al.(2009)Bickel, Ritov and Tsybakov}]{Ritov2009}
\bibinfo{author}{P.~J. Bickel}, \bibinfo{author}{Y.~Ritov},
  \bibinfo{author}{A.~B. Tsybakov}, \bibinfo{title}{Simultaneous analysis of
  lasso and {Dantzig selector}}, \bibinfo{journal}{Ann. Statist.}
  \bibinfo{volume}{37} (\bibinfo{year}{2009}) \bibinfo{pages}{1705--1732}.
%Type = Article
\bibitem[{Buldygin and Kozachenko(1980)}]{Buldygin1980}
\bibinfo{author}{V.~Buldygin}, \bibinfo{author}{Y.~Kozachenko},
  \bibinfo{title}{Subgaussian random variables}, \bibinfo{journal}{Ukrainian
  Mathematical Journal. Springer} \bibinfo{volume}{32} (\bibinfo{year}{1980})
  \bibinfo{pages}{483--489}.
%Type = Article
\bibitem[{Bunea et~al.(2011)Bunea, She and Wegkamp}]{bunea2011optimal}
\bibinfo{author}{F.~Bunea}, \bibinfo{author}{Y.~She},
  \bibinfo{author}{M.~Wegkamp}, \bibinfo{title}{Optimal selection of reduced
  rank estimators of high-dimensional matrices}, \bibinfo{journal}{Ann.
  Statist.} \bibinfo{volume}{39} (\bibinfo{year}{2011})
  \bibinfo{pages}{1282--1309}.
%Type = Article
\bibitem[{Bunea et~al.(2012)Bunea, She and Wegkamp}]{bunea2012joint}
\bibinfo{author}{F.~Bunea}, \bibinfo{author}{Y.~She},
  \bibinfo{author}{M.~Wegkamp}, \bibinfo{title}{Joint variable and rank
  selection for parsimonious estimation of high-dimensional matrices},
  \bibinfo{journal}{Ann. Statist.} \bibinfo{volume}{40} (\bibinfo{year}{2012})
  \bibinfo{pages}{2359--2388}.
%Type = Article
\bibitem[{Carroll et~al.(2006)Carroll, Ruppert, Stefanski and
  Crainiceanu}]{Carroll2010}
\bibinfo{author}{R.~J. Carroll}, \bibinfo{author}{D.~Ruppert},
  \bibinfo{author}{L.~A. Stefanski}, \bibinfo{author}{C.~M. Crainiceanu},
  \bibinfo{title}{Measurement error in nonlinear models}
  (\bibinfo{year}{2006}).
%Type = Article
\bibitem[{Datta and Zou(2017)}]{datta2017}
\bibinfo{author}{A.~Datta}, \bibinfo{author}{H.~Zou}, \bibinfo{title}{Cocolasso
  for high-dimensional error-in-variables regression}, \bibinfo{journal}{Ann.
  Statist.} \bibinfo{volume}{45} (\bibinfo{year}{2017})
  \bibinfo{pages}{2400--2426}.
%Type = Article
\bibitem[{Fan and Li(2001)}]{Fan2001}
\bibinfo{author}{J.~Fan}, \bibinfo{author}{R.~Li}, \bibinfo{title}{Variable
  selection via nonconcave penalized likelihood and its oracle properties},
  \bibinfo{journal}{J. Amer. Statist. Assoc.} \bibinfo{volume}{96}
  (\bibinfo{year}{2001}) \bibinfo{pages}{1348--1360}.
%Type = Article
\bibitem[{Izenman(2008)}]{Izenman2008}
\bibinfo{author}{A.~Izenman}, \bibinfo{title}{Modern multivariate statistical
  techniques: regression, classification, and manifold learning},
  \bibinfo{journal}{Springer, New York.}  (\bibinfo{year}{2008}).
%Type = Article
\bibitem[{Laurent and Massart(2000)}]{laurent2000adaptive}
\bibinfo{author}{B.~Laurent}, \bibinfo{author}{P.~Massart},
  \bibinfo{title}{Adaptive estimation of a quadratic functional by model
  selection}, \bibinfo{journal}{Ann. Statist.} \bibinfo{volume}{28}
  (\bibinfo{year}{2000}) \bibinfo{pages}{1302--1338}.
%Type = Article
\bibitem[{Li et~al.(2020)Li, Li and Ma}]{Li2020}
\bibinfo{author}{M.~Li}, \bibinfo{author}{R.~Li}, \bibinfo{author}{Y.~M. Ma},
  \bibinfo{title}{Inference in high-dimensional linear measurement error
  models}, \bibinfo{journal}{arXiv preprint}  (\bibinfo{year}{2020})
  \bibinfo{pages}{arXiv: 2001.10142}.
%Type = Article
\bibitem[{Liang and Li(2009)}]{liang2009variable}
\bibinfo{author}{H.~Liang}, \bibinfo{author}{R.~Li}, \bibinfo{title}{Variable
  selection for partially linear models with measurement errors},
  \bibinfo{journal}{Ann. Statist.} \bibinfo{volume}{104} (\bibinfo{year}{2009})
  \bibinfo{pages}{234--248}.
%Type = Article
\bibitem[{Liu et~al.(2015)Liu, Wang and Zhao}]{liu2015calibrated}
\bibinfo{author}{H.~Liu}, \bibinfo{author}{L.~Wang}, \bibinfo{author}{T.~Zhao},
  \bibinfo{title}{Calibrated multivariate regression with application to neural
  semantic basis discovery}, \bibinfo{journal}{J. Mach. Learn. Res.}
  \bibinfo{volume}{16} (\bibinfo{year}{2015}) \bibinfo{pages}{1579--1606}.
%Type = Article
\bibitem[{Loh and Wainwright(2012)}]{loh2011high}
\bibinfo{author}{P.~L. Loh}, \bibinfo{author}{M.~J. Wainwright},
  \bibinfo{title}{High-dimensional regression with noisy and missing data:
  {Provable guarantees with non-convexity}}, \bibinfo{journal}{Ann. Statist.}
  \bibinfo{volume}{40} (\bibinfo{year}{2012}) \bibinfo{pages}{1637--1664}.
%Type = Article
\bibitem[{Ma and Li(2010)}]{Lirunze2010}
\bibinfo{author}{Y.~Ma}, \bibinfo{author}{R.~Li}, \bibinfo{title}{Variable
  selection in measurement error models}, \bibinfo{journal}{Bernoulli}
  \bibinfo{volume}{16} (\bibinfo{year}{2010}) \bibinfo{pages}{274–300}.
%Type = Article
\bibitem[{Mishra et~al.(2017)Mishra, Dey and Chen}]{mishra2017sequential}
\bibinfo{author}{A.~Mishra}, \bibinfo{author}{D.~K. Dey},
  \bibinfo{author}{K.~Chen}, \bibinfo{title}{Sequential co-sparse factor
  regression}, \bibinfo{journal}{J. Comput. Graph. Statist.}
  \bibinfo{volume}{26} (\bibinfo{year}{2017}) \bibinfo{pages}{814--825}.
%Type = Article
\bibitem[{Rosenbaum and Tsybakov(2010)}]{rosenbaum2010sparse}
\bibinfo{author}{M.~Rosenbaum}, \bibinfo{author}{A.~Tsybakov},
  \bibinfo{title}{Sparse recovery under matrix uncertainty},
  \bibinfo{journal}{Ann. Statist.} \bibinfo{volume}{38} (\bibinfo{year}{2010})
  \bibinfo{pages}{2620--2651}.
%Type = Article
\bibitem[{St{\"a}dler and B{\"u}hlmann(2012)}]{stadler2012missing}
\bibinfo{author}{N.~St{\"a}dler}, \bibinfo{author}{P.~B{\"u}hlmann},
  \bibinfo{title}{Missing values: sparse inverse covariance estimation and an
  extension to sparse regression}, \bibinfo{journal}{Stat. Comput.}
  \bibinfo{volume}{22} (\bibinfo{year}{2012}) \bibinfo{pages}{219--235}.
%Type = Article
\bibitem[{Starbird and Palen(2012)}]{starbird2012will}
\bibinfo{author}{K.~Starbird}, \bibinfo{author}{L.~Palen},
  \bibinfo{title}{(how) will the revolution be retweeted?: information
  diffusion and the 2011 {Egyptian uprising}}, \bibinfo{journal}{Proceedings of
  the ACM 2012 Conference on Computer Supported Cooperative Work}
  (\bibinfo{year}{2012}) \bibinfo{pages}{7--16}.
%Type = Article
\bibitem[{Sun and Zhang(2012)}]{sun2012scaled}
\bibinfo{author}{T.~Sun}, \bibinfo{author}{C.-H. Zhang}, \bibinfo{title}{Scaled
  sparse linear regression}, \bibinfo{journal}{Biometrika} \bibinfo{volume}{99}
  (\bibinfo{year}{2012}) \bibinfo{pages}{879--898}.
%Type = Article
\bibitem[{Uematsu et~al.(2019)Uematsu, Fan, Chen, Lv and Lin}]{Uematsu17}
\bibinfo{author}{Y.~Uematsu}, \bibinfo{author}{Y.~Fan},
  \bibinfo{author}{K.~Chen}, \bibinfo{author}{J.~Lv}, \bibinfo{author}{W.~Lin},
  \bibinfo{title}{Sofar: large-scale association network learning},
  \bibinfo{journal}{IEEE Trans. Inform. Theory} \bibinfo{volume}{65}
  (\bibinfo{year}{2019}) \bibinfo{pages}{4924--4939}.
%Type = Article
\bibitem[{Yang and Lee(2017)}]{Kaiyang2017}
\bibinfo{author}{K.~Yang}, \bibinfo{author}{L.-F. Lee},
  \bibinfo{title}{Identification and {QML} estimation of multivariate and
  simultaneous equations spatial autoregressive models}, \bibinfo{journal}{J.
  Econometricss} \bibinfo{volume}{196} (\bibinfo{year}{2017})
  \bibinfo{pages}{196--214}.
%Type = Article
\bibitem[{Yang and Lee(2019)}]{Kaiyang2019}
\bibinfo{author}{K.~Yang}, \bibinfo{author}{L.-F. Lee},
  \bibinfo{title}{Identification and estimation of spatial dynamic panel
  simultaneous equations models}, \bibinfo{journal}{J. Econometrics}
  \bibinfo{volume}{76} (\bibinfo{year}{2019}) \bibinfo{pages}{32--46}.
%Type = Article
\bibitem[{Zheng et~al.(2019{\natexlab{a}})Zheng, Bahadori, Liu and
  Lv}]{bahadori2016scalable}
\bibinfo{author}{Z.~Zheng}, \bibinfo{author}{M.~T. Bahadori},
  \bibinfo{author}{Y.~Liu}, \bibinfo{author}{J.~Lv}, \bibinfo{title}{Scalable
  interpretable multi-response regression via {SEED}}, \bibinfo{journal}{J.
  Mach. Learn. Res.} \bibinfo{volume}{20} (\bibinfo{year}{2019}{\natexlab{a}})
  \bibinfo{pages}{1--34}.
%Type = Article
\bibitem[{Zheng et~al.(2018)Zheng, Li, Yu and Li}]{Zheng18}
\bibinfo{author}{Z.~Zheng}, \bibinfo{author}{Y.~Li}, \bibinfo{author}{C.~Yu},
  \bibinfo{author}{G.~Li}, \bibinfo{title}{Balanced estimation for
  high-dimensional measurement error models}, \bibinfo{journal}{Computational
  Statistics \& Data Analysis} \bibinfo{volume}{128} (\bibinfo{year}{2018})
  \bibinfo{pages}{78--91}.
%Type = Article
\bibitem[{Zheng et~al.(2019{\natexlab{b}})Zheng, Wu, Li and Wang}]{Zheng19}
\bibinfo{author}{Z.~Zheng}, \bibinfo{author}{J.~Wu}, \bibinfo{author}{Y.~Li},
  \bibinfo{author}{Y.~Wang}, \bibinfo{title}{Sequential scaled sparse factor
  regression}, \bibinfo{journal}{Manuscript}
  (\bibinfo{year}{2019}{\natexlab{b}}).
%Type = Article
\bibitem[{Zhu et~al.(2019)Zhu, Huang, Pan and Wang}]{zhu2019}
\bibinfo{author}{X.~Zhu}, \bibinfo{author}{D.~Huang}, \bibinfo{author}{R.~Pan},
  \bibinfo{author}{H.~Wang}, \bibinfo{title}{Multivariate spatial
  autoregressive model for large scale social networks}, \bibinfo{journal}{J.
  Econometrics}  (\bibinfo{year}{2019}).

\end{thebibliography}
\bibliographystyle{myjmva}

\end{document}